\documentclass[reprint,aps,pra,superscriptaddress,twocolumn,floatfix,amsmath,amssymb,amsfonts,longbibliography]{revtex4-2}
\usepackage{graphicx}
\usepackage[
colorlinks,
citecolor=blue,
urlcolor=blue,
linkcolor=red,breaklinks]{hyperref}
\usepackage[utf8]{inputenc}
\usepackage[vietnamese, english]{babel}
\usepackage{braket}
\usepackage{xcolor}
\usepackage{soul}
\renewcommand{\theequation}{S\arabic{equation}}
\renewcommand{\thefigure}{S\arabic{figure}}

\usepackage{physics}

\usepackage{blkarray}
\usepackage[dvipsnames]{xcolor}
\usepackage[utf8]{inputenc}
\usepackage{chngcntr}
\usepackage{xcolor}
\usepackage{braket}
\usepackage{sidecap}
\usepackage{lipsum}
\usepackage{bbold}
\usepackage{array}
\usepackage{comment}
\usepackage{amsmath,amssymb}
\usepackage{mathtools}
\usepackage{dsfont}

\begin{document}

\title{Interaction-Driven Charge Textures and Unconventional Superconductivity in Strained Monolayer Graphene}
\author{Elias Andrade}
\affiliation{
 Posgrado en Ciencias F\'{i}sicas, Universidad Nacional Aut\'{o}noma de M\'{e}xico (UNAM). Apdo. Postal 20-364, 01000 M\'{e}xico CDMX, Mexico
 }
 \affiliation{Department of Physics, University of Arkansas, Fayetteville, AR, USA}
\author{Alejandro Jimeno-Pozo}
\affiliation{Donostia International Physics Center, Paseo Manuel de Lardiz\'abal 4, 20018 San Sebastián, Spain}
\affiliation{IMDEA Nanoscience, Faraday 9, 28049 Madrid, Spain}
\author{Pierre A. Pantale\'on}
\email{panta2mi@gmail.com}
\altaffiliation[Current address: ]{Center for Theoretical Physics, Hainan University, Haikou 570228, China.}
\affiliation{IMDEA Nanoscience, Faraday 9, 28049 Madrid, Spain}
\affiliation{Department of Physics, Colorado State University, Fort Collins, Colorado 80523, USA}
\author{Francisco Guinea}
\affiliation{IMDEA Nanoscience, Faraday 9, 28049 Madrid, Spain}
\affiliation{Donostia International Physics Center, Paseo Manuel de Lardiz\'abal 4, 20018 San Sebastián, Spain}
\author{Gerardo G. Naumis}
\email{naumis@fisica.unam.mx}
\affiliation{
 Depto. de Sistemas Complejos, Instituto de F\'{i}sica, Universidad Nacional Aut\'{o}noma de M\'{e}xico (UNAM). Apdo. Postal 20-364, 01000 M\'{e}xico CDMX, Mexico
 }

\date{\today}

\begin{abstract}
Two-dimensional systems with flat bands support correlated phases such as superconductivity. While twisted moiré systems, like twisted bilayer graphene, have revealed such states, they remain complex to control. Here, we study monolayer graphene under uniaxial periodic strain, which forms a quasi-one-dimensional moir\'e lattice that hosts two flat sublattice-polarized bands. When electron-electron interactions are included at the self-consistent Hartree level, we find features familiar from moir\'e flat-band systems, such as Fermi-level pinning to the van Hove singularity and a Kohn--Luttinger-like pairing instability. In addition, we find strong interband enhancement of the pairing scale and higher-energy metastable electrostatic textures in enlarged supercells, with charge localized in selected regions of multiple unit cells.
\end{abstract}

\maketitle

\begin{figure}[ht]
\begin{center}
{\includegraphics[scale=1]{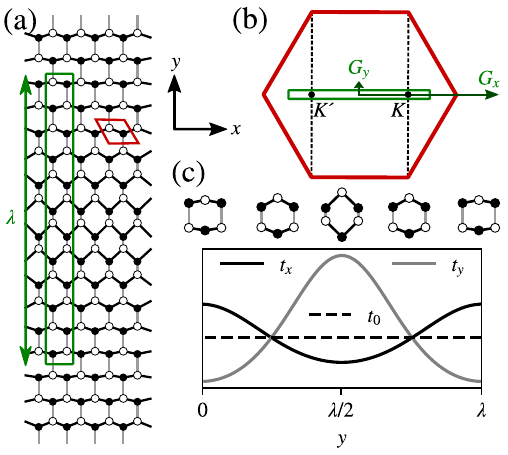}}
\caption{(a) Schematic of the system. An uniaxial strain is applied to a honeycomb lattice along the $y$-direction. A moir\'e pattern between the strain and the lattice produces a supercell of length $\lambda$. The unit cells for the honeycomb lattice and the supercell are shown in red and green, respectively. (b) The corresponding BZs for the unit cells shown in (a). For the supercell the BZ is folded into a quasi-1D mBZ with reciprocal lattice vectors $G_x$ and $G_y$. (c) Modulation of the two types of hoppings} $t_x$ (black) and $t_y$ (gray) along the $y$-direction. A schematic of the local deformation is shown above.
\label{fig:System}
\end{center}
\end{figure}
{\it Introduction.\textendash}
The discovery of superconductivity (SC) and other correlated phases in twisted bilayer graphene (TBG)~\cite{cao2018unconventional,cao2018correlated,yankowitz2019,lu2019superconductors,Xie2019Spectroscopic,sharpe2019emergent,Jiang2019Charge,choi2019electronic,stepanov2020untying,oh2021evidence} marked a turning point in the study of strongly correlated phenomena in two-dimensional materials. These findings confirmed that moir\'e patterns can create electronic flat bands, enhancing electron-electron interactions and enabling novel quantum phases~\cite{suarez2010,bistritzer2011moire}. Since then, a wealth of research has revealed rich phase diagrams in both twisted~\cite{park2021tunable,hao2021electric,kim2022evidence,shen2020correlated,cao2020tunable,park2022robust,zhang2022promotion,kuiri2022spontaneous,su2023superconductivity} and non-twisted~\cite{zhou2021superconductivity,zhou2022isospin,zhang2023enhanced,liu2024spontaneous,han2024correlated,holleis2025nematicity,han2025signatures} graphene systems. In particular, recent works have provided evidence for unconventional superconducting behavior in twisted moir\'e multilayers, although the microscopic pairing mechanism remains under active debate~\cite{stepanov2020untying,gao2024double,Park2025,barrier2024coulomb}.

Motivated by these findings, engineering flat bands has become a central goal in the field. One common route to flat bands involves applying magnetic fields to localize electrons into Landau orbits, supporting correlated phases like the fractional quantum Hall effect~\cite{Andrei1988observation,du2009fractional,bolotin2009observation}. However, this breaks time-reversal symmetry and requires strong fields. Strain-induced pseudo-magnetic fields offer a compelling alternative: they preserve time-reversal symmetry~\cite{levy2010strain,guinea2010energy,guinea2010generating,masir2013pseudo,Kauppila2016flat,jiang2017visualizing,naumis2017review,mao2020evidence,naumis2024review2,cao2020elastic,zhang2022strain,Yuan2024flat} and can reach effective field strengths of hundreds of Tesla~\cite{levy2010strain}. Furthermore, in some cases strain can be used to tune magnetic properties~\cite{ma2012Evidence,Guo2024Orbital}. Recent advances allow for diverse strain profiles, including origami folds~\cite{yang2022origami}, crenelated structures~\cite{kerjouan2024crenelated}, and strain superlattices~\cite{banerjee2020strain}. Theoretically, strain-modulated superlattices have been proposed as flat-band platforms analogous to twisted systems~\cite{timmel2020dirac,gao2023untwisting,wan2023nearly,meng2024flat}. A related route based on shear in bilayer graphene has also been proposed, where quasi-one-dimensional flat bands support a strong-coupling pairing scenario~\cite{gonzalez2026High}. This motivates asking whether charge reconstruction and pairing can arise in an even simpler single-layer setting.

In this Letter, we analyze monolayer graphene under uniaxial periodic strain, where lattice-strain mismatch creates a quasi-one-dimensional moir\'e pattern~\cite{Naumis2014Mapping,roman2017topological,Elias2023Topological,andrade2024flat}. This single-layer setup is analytically and numerically tractable. We show that the self-consistent Hartree potential produces more than a band renormalization: it yields one-cell sublattice-polarized solutions and higher-energy metastable charge textures in enlarged supercells. Using the reconstructed one-cell normal states, we further find a Kohn--Luttinger-like linearized pairing instability with strong interband enhancement.

{\it Model.\textendash}
\begin{figure*}[ht]
\begin{center}
{\includegraphics[width=1\textwidth]{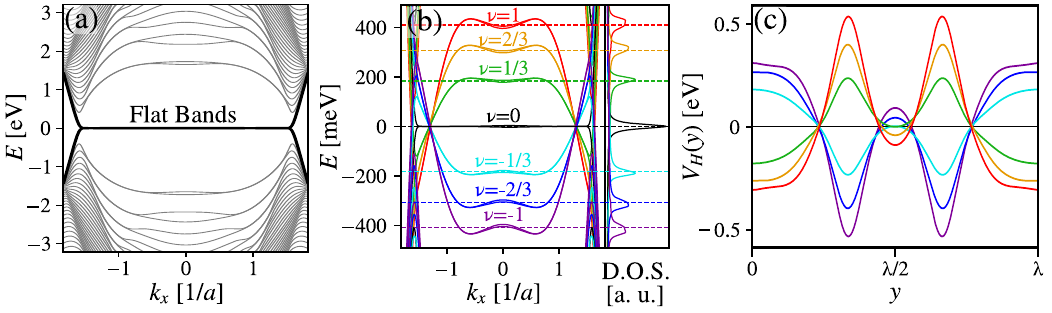}}
\caption{(a) Band structure of strained monolayer graphene with $A=0.15a$ and $\lambda=13.6$ nm where twofold-degenerate flat band (bold black) appear at the middle of the spectra. (b) Self-consistent Hartree bands and DOS with $\varepsilon=10$ for different filling fractions. The Fermi energy is indicated by a corresponding dashed line of the same color. (c) Real space Hartree potential.}
\label{fig:HartreeBands}
\end{center}
\end{figure*}
We start by considering a honeycomb lattice subjected to a periodic uniaxial strain along the $y$-direction. The strain modifies the atomic positions according to $\{x,y\}\rightarrow\{x,y+u(y)\}$, where $u(y)$ defines the strain profile. This profile is modeled using a sinusoidal function for the displacements, given by,
\begin{equation}
 u(y)=A\text{cos}\left(\frac{2 \pi y}{\lambda_0}+\phi\right),
 \label{Eq:DisFieldM}
\end{equation}
where $A$ is the amplitude of the deformation, $\lambda_0$ the wavelength, and $\phi$ a phase. Because the system has translational symmetry along the $x$-direction, the Hamiltonian can be mapped to an effective 1D bipartite lattice model (see Supplementary Material Fig.~S1~\cite{SM}).
We assume that the change in the distance between nearest neighbor sites translates into a modulation of the hoppings $t_x(y)$ and $t_y(y)$~\cite{AccurateTBStrain2018}.
If we consider a wavelength with a slight variation from the sublattice periodicity $\lambda_{sl}=3a/2$ with $a=0.142$ nm being the distance between carbon atoms, such that $\lambda_0^{-1}=\lambda_{sl}^{-1}+\lambda^{-1}$ with $\lambda>>\lambda_{sl}$, a moir\'e pattern appears with a moir\'e length given by $\lambda$ (see Supplementary Material Sec.~I~\cite{SM}). In Fig.~\ref{fig:System}(a) we show the deformed lattice, highlighting in green the supercell of length $\lambda$ along with the usual cell of the honeycomb lattice in red. In Fig.~\ref{fig:System}(b) we show the corresponding hexagonal Brillouin zone (BZ) and the folding into a quasi-1D rectangular mini-BZ (mBZ). The moir\'e pattern results in an out of phase modulation between $t_x$ and $t_y$ as shown in Fig.~\ref{fig:System}(c), the regions $t_x>t_y$ and $t_x<t_y$ correspond to different topological regions, as a result two soliton states appear around the domain walls $t_x=t_y$, each with opposite sublattice polarization~\cite{Elias2023Topological}. These almost localized states appear in the spectrum as flat bands at $E=0$ for a wide range of values of $k_x$, as shown in Fig.~\ref{fig:HartreeBands}(a). In the following, we will introduce the role of electronic correlations associated with these flat bands.

{\it Electrostatic Hartree Potential.\textendash}
We take into account the long-range Coulomb interaction by including an additional Hartree term. In twisted graphene systems, a Hartree correction leads to an electrostatic distortion larger than the bandwidth of the narrow bands and it strongly modifies the band structure close to the Fermi level~\cite{Guinea2018Electrostatic,Rademaker2018Charge,Cea2019Electronic,Cea2020band}. In our system, a similar effect is found. We perform a plane wave expansion of the Hamiltonian within the supercell (see Supplementary Material Sec.~III~\cite{SM}), so that the Hartree potential is also written as a superposition of plane waves in the strained direction,

\begin{equation}
 v_H(y)=\sum_n v_c(G_n)\delta\rho(G_n)e^{iG_ny},
 \label{Eq:HartreeReal}
 \end{equation}
where $G_n=nG_y$, $v_c(G_n)=\frac{e^2}{\varepsilon\sqrt{3}a|n|}$ the Fourier transform of the Coulomb potential, and the Fourier amplitudes of the charge density are given by~\cite{Guinea2018Electrostatic,Cea2019Electronic,Pantaleon2021Narrow},
\begin{small}
\begin{equation}
 \delta \rho(G_n)=2\int_{\text{mBZ}}\frac{d^2k}{V_{\text{mBZ}}}\sum_{i,l,m}\Delta f_{l,\boldsymbol{k}}(\nu)\phi_{l,k}^{i*}(G_m)\phi_{l,k}^{i}(G_m+G_n),
 \label{Eq:derhon}
\end{equation}
\end{small}
where $V_{\text{mBZ}}$ is the area of the mBZ, $l$ a band index resulting from the diagonalization of the full Hartree Hamiltonian, $H+V_H$, $\Delta f_{l,\boldsymbol{k}}(\nu)=f_{l,\boldsymbol{k}}(\nu)-f_{l,\boldsymbol{k}}(0)$ is the occupation measured with respect to charge neutrality, with the chemical potential fixed self-consistently to obtain the desired filling, and the factor 2 takes into account spin degeneracy. The Hartree potential in Eq.~\eqref{Eq:HartreeReal} implicitly depends on the fractional filling fraction $\nu$ of the conduction band, where $\nu=+2$ corresponds to fully filled valence and conduction bands and $\nu=-2$ corresponds to the case where both are empty. Note that small values of the dielectric constant induce a strong band distortion, which may cause the fully filled (or empty) middle bands to mix with nearby high-energy bands. In the following, we consider how interactions modify the electronic structure and can produce self-consistent Hartree states that can be classified by two independent properties: their sublattice character, either {\it sublattice symmetric} or {\it sublattice polarized}, and their translational periodicity, either repeating after one moir\'e unit cell or only after an enlarged moir\'e supercell.

 {\it Fermi-level Pinning.\textendash}
In Fig.~\ref{fig:HartreeBands}(b) we show the Hartree bands as a function of the filling.\ Interestingly, the flat bands exhibit a strong dependence on filling.\ The maximum amplitude of the Hartree potential fluctuations is approximately $\Delta V_H \simeq 500$ meV for $|\nu| = 1$ and $\varepsilon = 10$, which is significantly larger than the bandwidth of the flat bands. Notably, a similar effect is obtained in TBG near the magic angle; however, under the same dielectric environment, the resulting band distortion is only about $20$ meV, an order of magnitude smaller~\cite{Guinea2018Electrostatic,Rademaker2018Charge,Cea2019Electronic}.\ For $|k_x| \lesssim 1/a$, the wavefunctions are centered at the domain walls where the magnitudes of $t_x$ and $t_y$ are inverted and they are notably similar to one another (see Supplementary Material Fig.~S2~\cite{SM}), whereas at the region with $|k_x| \gtrsim 1/a$, the states closely resemble those in the remote bands. Because the Hartree potential is the charge density contribution from charge neutrality to a given filling, states with charge distributions similar to that of the Hartree potential exhibit a strong response to it, while states with differing charge distributions are less affected~\cite{Pantaleon2021Narrow}. Consequently, there is a pinning of the vHs to the Fermi energy.\ This remarkable result is analogous to what has been obtained in TBG~\cite{Guinea2018Electrostatic,Cea2019Electronic}. The quenching of kinetic energy due to the non-dispersive flat bands enhances the importance of electronic interactions, particularly when the Fermi energy is close to the vHs, as found in our system. Notably, as shown in Fig.~\ref{fig:HartreeBands}(c), the Hartree potential exhibits a significant magnitude even for a moderate dielectric constant. This strong Hartree response indicates a nontrivial charge distribution and a substantial Coulomb-driven reconstruction of the flat bands. In the present system, this reconstructed normal state is the starting point for the pairing analysis below. A strong filling-dependent modification of the band shape may also be viewed as a signature of an emergent interaction beyond a fixed single-particle band description, often discussed in terms of electron-assisted hopping~\cite{Hirsch1990Superconductivity,Marsiglio1990Hole}. Such interaction-generated hopping processes have been proposed as a possible ingredient for superconductivity in other correlated systems~\cite{Guinea2018Electrostatic}. A more detailed mapping of the Hartree potential is provided in Supplementary Material Sec.~IV~\cite{SM}.

\begin{figure*}[ht]
\begin{center}
{\includegraphics[width=1\textwidth]{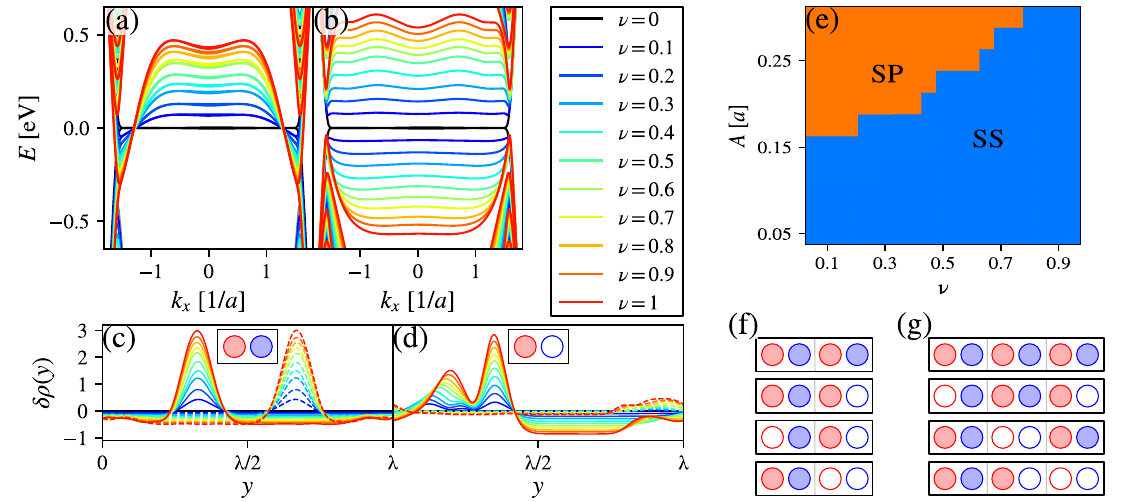}}
\caption{Comparison between the two self-consistent Hartree solutions found for $A=0.15a$, $\lambda=13.6$ nm and $\varepsilon=10$. Band structure at different fractional fillings for the (a) SS solutions and (b) SP solutions. (c)-(d) Show the inhomogeneity in the charge distribution for both SS and SP solutions respectively. The solid lines show the contributions from sublattice A while the dashed lines for sublattice B. Similarly, the insets show a schematic of the localization, here red and blue circles represent sublattice A and B, respectively. (e) Phase diagram as a function of filling fraction and strain amplitude, here $\lambda=27.2$ nm and $\varepsilon=6$ were used. (f)-(g) Schematic of the localization for solutions within multiple unit cells. The top rows in (f) and (g) show one-cell periodic solutions represented in enlarged supercells, and are therefore equivalent to the corresponding one-cell solutions after band unfolding. The lower rows show metastable textures whose charge distribution is not equivalent to a simple repetition of the one-cell solution and therefore has reduced translational symmetry. In (f) two unit cells are considered with $\nu=1/2$, while in (g) three unit cells with $\nu=1/3$.}
\label{fig:Solutions}
\end{center}
\end{figure*}

{\it Electrostatically Induced Polarized States\textendash}
The self-consistent solutions of the Hartree Hamiltonian in Fig.~\ref{fig:HartreeBands} are symmetric with respect to the half of the unit cell; these one-cell {\it sublattice symmetric} (SS) states are the usual ground states for a given filling. However, we find additional one-cell states with similar energy that strongly depend on the initial condition in the self-consistent calculation. These states are gapped and sublattice polarized, and we identify them as {\it sublattice polarized} (SP) states. Thus, the SP solution is an electrostatic Hartree state of the strained monolayer, rather than a spin- or exchange-driven ordered phase. In Figs.~\ref{fig:Solutions}(a)-(b), we show the band structures for the SS and SP states, respectively, for filling fractions between $\nu=0$ and $\nu=1$. Similarly to the SS states, the flat band regions of the SP states undergo a rigid energy shift; however, in this case, the bands shift in opposite directions, lifting the degeneracy. For each filling considered, the charge distributions for both SS and SP states are shown in Figs.~\ref{fig:Solutions}(c)-(d), respectively. It can be seen that, in the SP states, the charge is mostly distributed on one sublattice. This sublattice-symmetry breaking is the origin of the gap seen in the bands. Notably, we find that SP solutions can become the ground state at small $\nu$ if both the interaction strength and the localization are enhanced. As shown in the phase diagram in Fig.~\ref{fig:Solutions}(e), the former is achieved by reducing the dielectric constant, while the latter is realized by increasing the amplitude of the strain. We note that higher values of $A$ may be complicated to realize experimentally, but as the ratio $A/\lambda_0$ controls the band flatness, similar results may be obtained for smaller $A$ with a larger $\lambda_0$. This should not be interpreted as homogeneous tensile loading of pristine graphene. The present model describes a nonuniform periodic strain-engineered hopping modulation, for which the relevant control parameter is the spatial modulation of the bond lengths and hoppings. Graphene strain engineering has demonstrated large elastic deformations and controlled nonuniform strain profiles~\cite{Lee2008Measurement,Si2016Strain,Tomori2011Introducing}. In addition, in Supplementary Material Sec.~XI~\cite{SM} we included the effects of the exchange term and verified that the ground state corresponds to the non-spin-polarized case.

{\it Interaction-Driven Charge Textures.\textendash}
As mentioned previously, for a single unit cell, shown in Figs.~\ref{fig:Solutions}(c) and (d), the SS phase exhibits identical charge distribution across both sublattices.\ In contrast, in the SP phase, inversion symmetry is broken, and the charge is predominantly localized on one sublattice.\ This behavior has important implications when extending our analysis to multiple unit cells. Figures~\ref{fig:Solutions}(f) and (g) show schematic representations of charge localization in systems with two and three unit cells.\ The top rows of each panel illustrate the SS phase as obtained in an inversion-symmetric system.\ However, we also find metastable solutions that break inversion and translational symmetry (see also Supplementary Material Figs.~S6 and S7 and Tables~II and III~\cite{SM}). Here metastable refers to converged self-consistent Hartree solutions that remain stable under iteration for enlarged-cell seeds. Tables~II and III in Supplementary Material~\cite{SM} show that, for the two-cell case at $\nu=1/2$ and the three-cell case at $\nu=1/3$, the lowest-energy state among the converged solutions preserves the one-cell periodicity. The reduced-periodicity textures are therefore higher-energy metastable states, and the superconducting calculation below uses the one-cell SS and SP parent states.

 These configurations arise from electrostatic interactions and lead to charge textures with different spatial arrangements. Importantly, only a fraction of each unit cell becomes strongly occupied, while the remaining regions remain weakly populated. These broken-symmetry states may be compared only at a phenomenological level with Tao--Thouless-like charge patterns~\cite{Tao1983Fractional}: in both cases, the density forms enlarged-period textures with weight concentrated in selected regions of the enlarged cell. We do not claim quantized fractional charge or an adiabatic connection to fractional quantum Hall states~\cite{Seidel2006Abelian,Bergholtz2008Quantum}.

The sensitivity to initial conditions also suggests that weak sample inhomogeneity could favor such reduced-periodicity configurations. These higher-energy metastable textures are not the lowest-energy states among the converged solutions compared here, but they show that long-range Hartree interactions can generate enlarged-cell charge order in this minimal system. 

The enlarged-cell textures should therefore be understood as higher-energy metastable Hartree solutions for the parameters analyzed here, rather than as the thermodynamic ground state. Their significance is that they reveal an electrostatic route by which flat-band Coulomb interactions can generate reduced-periodicity charge configurations. Recent STM experiments in bilayer graphene aligned with hBN have directly visualized interaction-driven Chern-insulator phases that double, triple, or quadruple the moir\'e unit cell~\cite{He2026Visualizing}, showing that related graphene flat-band systems can support enlarged-cell electronic reconstructions.

{\it Superconductivity.\textendash} We now ask whether the reconstructed one-cell normal states support an electronic pairing instability. We use a Kohn--Luttinger-like framework, where screening of the repulsive Coulomb interaction generates an attractive component in the Cooper channel~\cite{Kohn1965NewMechanism,Chubukov1993KohnLuttinger}. The Hartree potential fixes the normal-state bands, while the pairing kernel is built from the screened interaction projected into the Cooper channel. No Hartree term is reintroduced in the pairing kernel, and the OP is not fed back into the Hartree loop. The calculation is detailed in Supplementary Material Secs.~VII and VIII~\cite{SM}.

The pairing instability is assessed through the linearized gap equation:
\begin{equation}
 \Delta_{l_1,l_2}(\boldsymbol{k})=\sum_{\boldsymbol{k'}}\sum_{l_1',l_2'}\Gamma_{l_1,l_2}^{l_1',l_2'}(\boldsymbol{k,k'}) \Delta_{l_1',l_2'}(\boldsymbol{k'}),
\end{equation}
where $l_i$ and $l^{'}_i$, with $i={1,2}$, are band indices. Note that the above equation allows us to distinguish the intraband ($\Delta_{1,1}$ and $\Delta_{2,2}$) and interband ($\Delta_{1,2}$=$\Delta_{2,1}$) contributions to the pairing state. The critical scale $T_C$ is defined as the temperature at which the leading eigenvalue of the kernel reaches unity. It should be understood as the linearized instability scale of the reconstructed normal state~\footnote{The scale $T_C$ is defined as the temperature at which the pairing instability sets in. Owing to the strong anisotropy of the system, the actual superconducting transition may occur at a lower temperature due to phase fluctuations.}

\begin{figure}[ht]
\begin{center}
{\includegraphics[scale = 1]{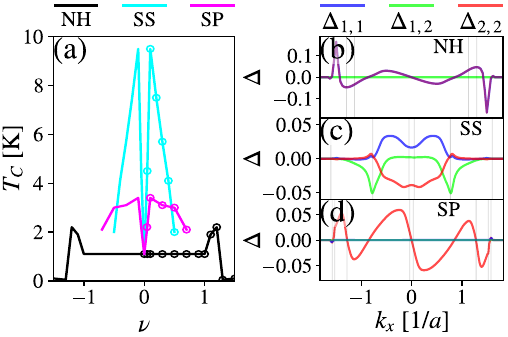}}
\caption{(a) Pairing scale as a function of fractional filling with $A=0.15a$, $\lambda=13.6$ nm and $\varepsilon=10$. The three cases are shown, NH (black), SS (cyan) and SP (magenta). The OPs are shown at $\nu=0.1$ for (b) NH, (c) SS and (d) SP.}
\label{fig:TC}
\end{center}
\end{figure}

Figure~\ref{fig:TC}(a) shows the calculated $T_C$ as a function of filling $\nu$ for three cases: the system without Hartree potential (NH), and with Hartree for both the SS and SP solutions. In the NH case, $T_C$ remains nearly constant around 1.1 K as long as the filling stays within the flat band region. Interestingly, the highest $T_C$ in this case is found near $\nu \approx \pm 1.2$, where the bands begin to disperse, beyond which $T_C$ decreases. When the Hartree potential is included, the pairing scale $T_C$ is significantly enhanced for both the SS and SP phases. In particular, for the SS phase, $T_C$ reaches 9.5 K at $\nu = 0.1$, while for the SP phase, it peaks at 3.4 K at a similar filling.

In Figs.~\ref{fig:TC}(b)–(d), we display the corresponding OP at $\nu = 0.1$. For each case, both intraband and interband contributions are shown. The gray lines indicate the positions of the Fermi surface~\footnote{Note that, in this quasi-one-dimensional system, the Fermi surface consists of stripes with a small width in $k_y$.}. It is evident that the OP is not strictly localized at the Fermi surface but instead has significant components across a range of momenta. This depinning behavior resembles that observed in graphene multilayers~\cite{ParraMartinez2025QuarterMetal,Qin2024Chiral} and TMDs~\cite{Chen2026Finite} and results from the presence of multiple states at the Fermi energy.

As the OP corresponds to the wave function of the electron pair, it must be antisymmetric under particle exchange. Within the spin-compensated parent states studied here, the SS case has an even-parity OP, compatible with spin-singlet pairing, while the NH and SP cases have odd-parity OPs, compatible with spin-triplet pairing. This classification does not directly apply to spin-polarized parent states, where the allowed channels must be analyzed separately. Our calculation uses the full folded lattice basis, not a valley-projected continuum model. The low-energy flat-band wave functions therefore contain folded components associated with both graphene valleys, and valley is not used as an independent conserved flavor in the present pairing kernel.

 Parity is not the only feature distinguishing the SS case. As shown in Fig.~\ref{fig:TC}(c), it is the only scenario with a nonzero interband OP (green line). This behavior may be linked to the band degeneracy at the Fermi surface, a feature absent in the NH and SP cases. Notably, we find that interband contributions play a critical role in determining this instability scale. When these contributions are neglected (see Supplementary Material, Sec.~X~\cite{SM}), the SS value of $T_C$ drops significantly, from 9.5 K to 2.2 K. Additionally, in Supplementary Material Sec.~XII~\cite{SM} the $T_C$ was calculated for other values of $A$, showing that the instability persists with scales of a few kelvin, even as the strain amplitude is reduced. This indicates that precise fine-tuning of this parameter is not necessary.

{\it Conclusions.\textendash} The microscopic origin of superconductivity in graphene-based moir\'e systems remains open, with several theoretical mechanisms under discussion. Our results show that periodically strained monolayer graphene provides a single-layer platform where long-range Hartree reconstruction and screened repulsive interactions can generate sizable superconducting instabilities.
This setup generates a one-dimensional moir\'{e} pattern that hosts flat bands with strong sublattice polarization and almost localized charge distributions.
Our main findings show that long-range Hartree electrostatics has a central role in this strained system: it pins the van Hove singularity to the Fermi level, reconstructs the flat bands, and generates both one-cell sublattice-polarized states and metastable enlarged-cell charge textures. The enlarged-cell textures are higher-energy metastable states in the cases analyzed here, while the one-cell periodic solution has the lowest energy among the converged states compared in Supplementary Material Tables~II and III. We further find that the reconstructed one-cell normal states support a Kohn--Luttinger-like instability from a screened repulsive Coulomb interaction, with the corresponding scale enhanced by interband contributions. The pairing strength also correlates with the underlying quantum geometry of the bands, as discussed in the SM Sec.~IX~\cite{SM}. This regularized quantum-metric analysis is used only as a diagnostic of band-geometric structure near the flat-band manifold, whereas the pairing enhancement itself is established directly from the linearized superconducting kernel. Circular-photogalvanic measurements in magic-angle twisted bilayer graphene further reveal reduced $C_1$ symmetry and Berry-curvature-dipole physics~\cite{Persky2026Revealing}, placing this diagnostic within the broader phenomenology of symmetry-reconstructed graphene moir\'e flat bands.

Our calculations give pairing scales of a few kelvin for experimentally motivated strained graphene parameters, for example through periodic ripples. The calculated $T_C$ should therefore be interpreted as the scale at which the linearized pairing instability appears, rather than as a direct estimate of the thermodynamic transition temperature. Strain profiles with amplitudes exceeding 10\% have already been observed~\cite{banerjee2020strain}. Smaller strain magnitudes, which are more commonly achieved experimentally~\cite{jiang2017visualizing,cao2020elastic,zhang2022strain}, should also support pairing instabilities, albeit with lower $T_C$ values, as shown in Supplementary Material Sec.~XII~\cite{SM}. We also find broken-symmetry textures induced by purely electrostatic interactions that could be probed through STS sublattice-resolved maps, while the Fermi-level pinning may be verified using compressibility measurements. Finally, phase-sensitive measurements, such as Josephson interferometry or SQUID-like junction geometries, could help test sign changes of the superconducting order parameter. Together with recent experiments on enlarged-cell Chern-insulator phases and symmetry-reduced quantum-geometric responses~\cite{He2026Visualizing,Persky2026Revealing}, these results support the use of periodically strained monolayer graphene as a minimal platform for isolating mechanisms that also operate in related graphene moir\'e flat bands.
These findings suggest that periodically strained monolayer graphene is not merely a simplified analogue of moiré materials, but a clean setting where electrostatic reconstruction, charge texture formation, and unconventional pairing can be studied at their common origin.

{\it Acknowledgments.\textendash} We thank Saul A. Herrera, Zhen Zhan, \foreignlanguage{vietnamese}{Võ Tiến Phong}, Ramon Carrillo-Bastos and Danna Liu for useful discussions.\ E.A.\ and G.G.N acknowledge financial support from UNAM DGAPA Project No.\ IN101924.\ P.A.P acknowledges funding by Grant No.\ JSF-24-05-0002 of the Julian Schwinger Foundation for Physics Research.\ IMDEA Nanociencia acknowledges support from the ‘Severo Ochoa’ Programme for Centres of Excellence in R\&D (CEX2020-001039-S/AEI/10.13039/501100011033). P.A.P. and F.G. acknowledge support from NOVMOMAT, project PID2022-142162NB-I00 funded by MICIU/AEI/10.13039/501100011033 and by FEDER, UE as well as financial support through the (MAD2D-CM)-MRR MATERIALES AVANZADOS-IMDEA-NC. A.J.-P. and F.G. acknowledge funding from the EU NextGenerationEU/PRTR-C17.I1, as well as by the IKUR Strategy under the collaboration agreement between Ikerbasque Foundation and DIPC on behalf of the Department of Education of the Basque Government. P.A.P thanks the hosting of the Department of Physics, Colorado State University in Fort Collins, where part of this project was developed. 


%

\clearpage

\onecolumngrid

\setcounter{equation}{0}
\setcounter{figure}{0}
\setcounter{table}{0}
\setcounter{page}{1}
\setcounter{section}{0}
\makeatletter
\renewcommand{\theequation}{S\arabic{equation}}
\renewcommand{\thefigure}{S\arabic{figure}}

\begin{center}
\Large {\it Supplementary Material for} \\
Interaction-Driven Charge Textures and Unconventional Superconductivity in Strained Monolayer Graphene
\end{center}
\begin{center}
\normalsize{Elias Andrade, Alejandro Jimeno-Pozo, Pierre A. Pantale\'on*, Francisco Guinea and Gerardo Naumis**}
\end{center}
\begin{center}
    \small {*panta2mi@gmail.com\\
    **naumis@fisica.unam.mx}
\end{center}
\tableofcontents

\section{Real Space Lattice Representation}
\label{Sec:SM1}
\begin{figure*}[ht]
\begin{center}
{\includegraphics[width=0.5\textwidth]{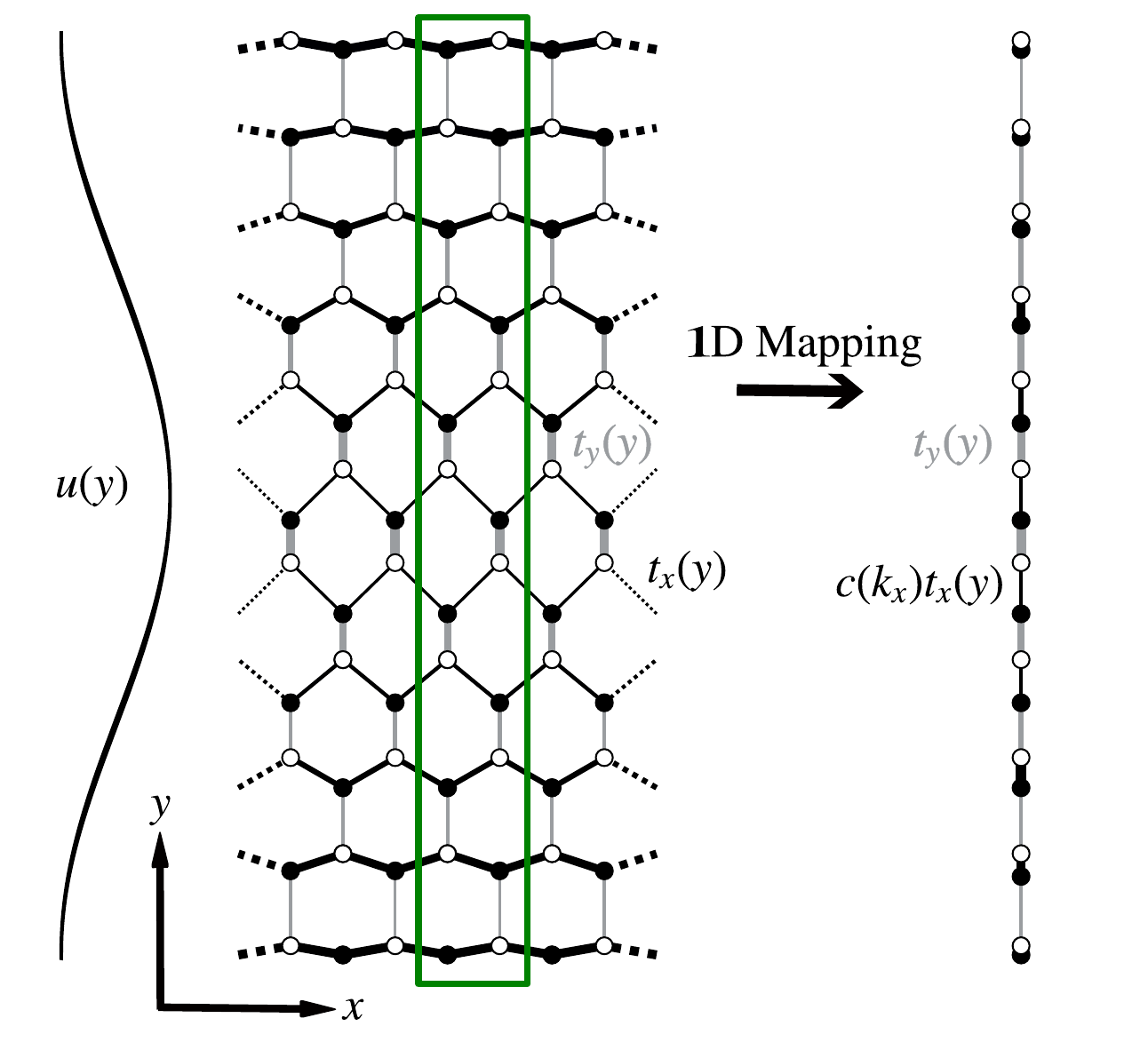}}
\caption{Schematics of the mapping from the lattice system to an effective 1D chain. The effective strain profile $u(y)$ is represented on the left. The green rectangle is the lattice unit cell.}
\label{fig:SMReal}
\end{center}
\end{figure*}

We consider a honeycomb lattice subject to an uniaxial strain along the $y$-direction. In particular, a sinusoidal strain is considered, such that the lattice sites displacement field is given by,
\begin{equation}
    u(y)=A\text{cos}\left(\frac{2 \pi y}{\lambda_0}+\phi\right),
    \label{Eq:DisField}
\end{equation}
where $A$ is the amplitude of the deformation and $\lambda_0$ the wavelength. As shown in Fig.~1, the lattice system has periodicity in the $x$-direction, and therefore, the Hamiltonian is given by~\cite{Naumis2014Mapping},
\begin{equation}
    H(k_x)=-t_0\sum_{n}[c(k_x)t_x(y_n)b_n^{\dagger}a_n+t_y(y_n)a_{n+1}^{\dagger}b_n]+\text{h.c.},
    \label{Eq:HamReal}
\end{equation}
where $a_n$ ($a_n^\dagger$) and $b_n$ ($b_n^\dagger$) are annihilation (creation) operators for the $n$-th site along the $y$-direction for sublattice A and B, respectively. The coefficient $c(k_x)=2\text{cos}(\sqrt{3}k_xa/2)$. As the atomic positions are now given by $(x',y')=(x,y+u(y))$, the hopping integrals acquire a $y-$dependence. In terms of the atomic displacement positions the modulated hoppings are given by,
\begin{subequations}
    \begin{equation}
        t_x(y)=t_0\text{exp}\left(-\frac{\beta}{2a}\left[u^A\left(y+\frac{3a}{2}\right)-u^B(y)\right] \right),
    \end{equation}
    \begin{equation}
        t_y(y)=t_0\text{exp}\left(-\frac{\beta}{a}[u^B(y)-u^A(y)] \right),
    \end{equation}
    \label{Eq:Modulation}
\end{subequations}
where $t_0=2.8$ eV is the nearest neighbor hopping integral of graphene, $\beta\approx3$ is the Gr\"uneisen parameter~\cite{AccurateTBStrain2018} and $u^{A/B}(y)$ is the value of the displacement field for sublattice A/B at a given coordinate $y$. An interesting feature of this model is the presence of flat bands arising from a moiré pattern created by a mismatch between the lattice period of the honeycomb structure and the wavelength of the lattice displacement field. We write the periodicity of such field $\lambda_0$ as,
\begin{equation}
    \frac{1}{\lambda_0}=\frac{1}{\lambda_{sl}}+\frac{1}{\lambda},
\end{equation}
where $\lambda_{sl}=3a/2$ is the periodicity between atoms of the same sublattice in the $y$-direction and $\lambda$ is the effective wavelength, giving the length of the moir\'e supercell. This can be shown explicitly by the substitution of the discrete positions of the sites in each sublattice $y^A_n=n\lambda_{sl}$ and $y^B_n=n\lambda_{sl}+a/2$ in Eq.~\eqref{Eq:DisField}. The effective displacement field seen by each sublattice is then given by,
\begin{subequations}
    \begin{equation}
        u^A(y)=A\text{cos}\left (\frac{2\pi y}{\lambda}+\phi \right),
    \end{equation}
    \begin{equation}
        u^B(y)=A\text{cos}\left (\frac{2\pi y}{\lambda}+\frac{2 \pi}{3}+\phi \right),
    \end{equation}
\end{subequations}
thus the displacement fields oscillate with a wavelength $\lambda$, but are out of phase by $\frac{2\pi}{3}$. As the hoppings in Eq.~\eqref{Eq:Modulation} depend on the difference of the displacements between neighboring sites, they can be rewritten as,
\begin{equation}
    t_\eta(y)=t_0\text{exp}\left[\alpha_{\eta} \text{sin}\left(\frac{2 \pi y}{\lambda}+\phi\right) \right],
    \label{Eq:SMHopping}
\end{equation}
where $\eta=x,y$, $\alpha_x=\sqrt{3}A \beta/2a$ and $\alpha_y=-\sqrt{3}A \beta/a$. Here we used the fact that $\lambda>>a$, thus $u^A\left(y+\frac{3a}{2}\right) \approx u^A(y)$. The modulation of the hoppings is shown schematically in Fig.~1(c) and the mapping from the lattice to the 1D system is shown in Fig.~\ref{fig:SMReal}.

\begin{figure*}[ht]
\begin{center}
{\includegraphics[width=1\textwidth]{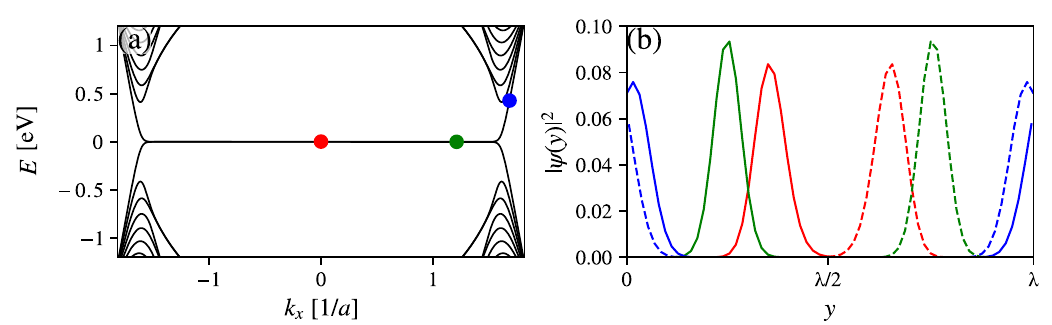}}
\caption{Electronic band structure and charge distribution of strained monolayer graphene obtained through direct diagonalization of Eq. (\ref{Eq:HamReal}) with $A=0.15a$ and $\lambda=13.6$ nm. (a) The band structure around the double degenerate flat band. The three color markers indicate the states whose probability density is shown in (b). The states of sublattice A (B) are shown by the solid (dashed) lines. The states present the highest localization near $aK_D=2\pi/3\sqrt{3}\approx1.209$ (green circle).
}
\label{fig:SMChargeLattice}
\end{center}
\end{figure*}

The modulation creates two topologically distinct regions: one where $t_x>t_y$ and the other where $t_x<t_y$. At the domain walls between these regions, the wavefunction becomes almost localized, resulting in flat bands in the spectrum. In Fig.~\ref{fig:SMChargeLattice} (a) we show the spectrum, highlighting three points at which the probability density is presented in Fig.~\ref{fig:SMChargeLattice} (b). At the momentum corresponding to the Dirac point of pristine graphene (green circle) the wavefunction is maximally localized. As we deviate from this value, the localization center shifts in real space. As $k_x$ approaches the BZ borders the state is lifted from zero energy towards the bulk states.

 It is easy to prove that the corresponding zero mode solutions are topological solitons localized near the domain walls. Consider the  wave function at the origin of the effective chain seen in Fig. \ref{Sec:SM1}, to which we assign the initial value $\psi_0^{A}$ on the $A$ sublattice and $\psi_0^{B}$ on the $B$ sublattice. Using the effective one-dimensional tight-binding equation for $E=0$, and since the displacement field changes slowly within the same sublattice, we can consider $u_{y_{n+1}}^{A/B} \approx u_{y_n}^{A/B}$ where $y_n$ is the strain at site $n$ in the chain seen in Fig. \ref{Sec:SM1}. This allows us to obtain the wave function at site $n$ in the chain, 
\begin{equation}
    \psi_n^{A/B}=[-c(k_x)]^{\pm n}\text{exp}\left[\pm \frac{3\beta}{2a}\sum_{j=0}^n \Delta u(y_j)\right]\psi_0^{A/B},
    \label{Eq:DiscWF}
\end{equation}
where $\Delta u(y_j)=u(y_j)^A-u(y_j)^B$. Whenever $\Delta u(y_j)$ is positive (negative), the wave function will grow (decay) for sublattice A, while the opposite happens for sublattice B. It is possible to write a continuum version of the previous equation,
\begin{equation}
    \psi^{A/B}(y,q_x)=N \text{exp}\left[\pm \int_y m(y',q_x)dy'\right],
    \label{Eq:wf}
\end{equation}
where $N$ is a normalization constant and $m(y',q_x)$  is a space dependent mass given by the pseudo-magnetic field produced by the strain \cite{Elias2023Topological,andrade2024flat}. Another way to produce such solutions is to show that the Hamiltonian Eq. (\ref{Eq:HamReal}) can be written as a Jackiw-Rebbi model, and therefore, Eq. (\ref{Eq:wf}) are the corresponding zero mode solutions of such model.

\section{Effects of including the Poisson Ratio}
The Poisson ratio $\nu_{P}$ accounts for the deformation induced perpendicular to the direction of the applied strain. In this case, it produces a slight modification of the $x$ components of the atomic positions, such that $\{x,y\}\rightarrow\{x-\nu_{P}u(y),y+u(y)\}$. We take $\nu_{P}=0.165$ for graphene~\cite{AccurateTBStrain2018}. Since the $t_y$ bonds are aligned along the $y$-axis, they remain unchanged, whereas the $t_x$ bonds are slightly modified as $\alpha_x \rightarrow \alpha_x(1-\sqrt{3}\nu_P)$.

\begin{figure}[ht]
\begin{center}
{\includegraphics[width=0.9\textwidth]{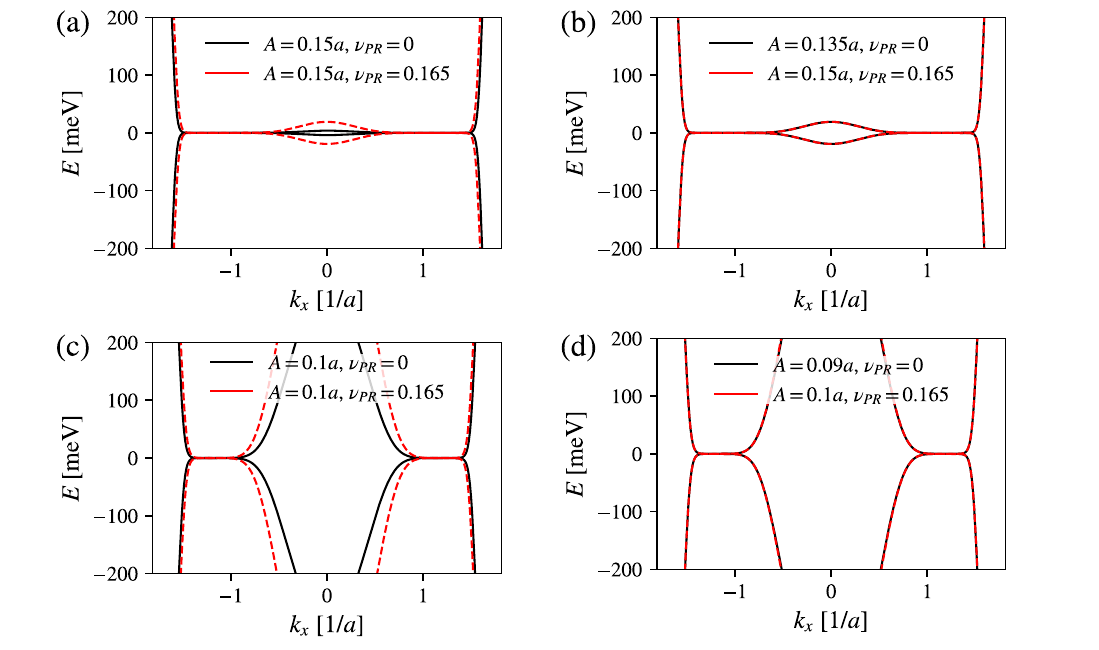}}
\caption{
Effects of the Poisson ratio in the band structure. (a) Comparison between the bands without ($\nu_{PR}=0$) and with ($\nu_{PR}=0.165$) Poisson ratio for $A=0.15a$. (b) Comparison between the inclusion of Poisson ratio and a $10\%$ decrease of the strain amplitude. (c)-(d) Similarly as in (a)-(b) with $A=0.1a$.}
\label{fig:SMPoisson}
\end{center}
\end{figure}

In Fig.~\ref{fig:SMPoisson}(a), we compare the band structures with and without including $\nu_{P}$ for $A=0.15a$. The effect is similar to that of reducing the strain amplitude. In particular, Fig.~\ref{fig:SMPoisson}(b) shows that reducing the strain amplitude by 10\% produces an equivalent spectrum to that obtained by including the Poisson ratio. Figs.~\ref{fig:SMPoisson}(c)-(d) analogously show the case for $A=0.1a$. Since the inclusion of the Poisson ratio does not qualitatively alter the results, we neglect it in the following analysis.

\section{Momentum Space Lattice Representation}
To better illustrate the coupling due to the modulation, we perform a Fourier transform of the modulated hoppings as they are periodic within the supercell,
\begin{equation}
    t_\eta(y) = t_0\sum_{m} f_{\eta,m} e^{iG_my}.
\end{equation}
Since $\alpha_\eta<1$, $t_\eta$ from Eq. (\ref{Eq:SMHopping}) is expanded up to second order, 
\begin{equation}
    \frac{t_\eta(y)}{t_0}=1+\alpha_\eta \text{sin}\left(\frac{2 \pi y}{\lambda}+\phi\right)
    +\frac{\alpha_\eta^2}{4}\left[ 1-\text{cos}\left(2\frac{2 \pi y}{\lambda}+2\phi\right) \right]+\mathcal{O}(\alpha_\eta^3),
\end{equation}
and from here the Fourier coefficients can be directly identified as, 
\begin{equation}
    \frac{f_{\eta,0}}{t_0}=1+\frac{\alpha_\eta^2}{4},
    \quad \frac{f_{\eta,1}}{t_0}=\frac{-i \alpha_\eta}{2}e^{i\phi}, \quad \frac{f_{\eta,2}}{t_0}=\frac{-\alpha_\eta^2}{8}e^{2i\phi}.
\end{equation}
The Fourier amplitudes satisfy,  $f_{\eta,-m}=f_{\eta,m}^*$, $\boldsymbol{G_m}=m\boldsymbol{G_y}$ and $G_m=|\boldsymbol{G_m}|=2\pi m/\lambda$. Because the system is periodic in the $x$-direction and strain modulated in the $y$-direction, we write the reciprocal lattice vectors as
\begin{equation}
    \boldsymbol{G_x}=\frac{2 \pi}{\sqrt{3}a}\{1,0\}, \quad \boldsymbol{G_y}=\frac{2 \pi}{\lambda}\{0,1\}.
\end{equation}

Now our Hamiltonian in Eq. (\ref{Eq:HamReal}) can be transformed into momentum space as,
\begin{equation}
    \mathcal{H}=\sum_{\boldsymbol{k}}\sum_{n}^{N_k} \sum_{m=-2}^2 h_m(\boldsymbol{k}+\boldsymbol{G_n})
a_{\boldsymbol{k}+\boldsymbol{G_n}+\boldsymbol{G_m}}^{\dagger}b_{\boldsymbol{k}+\boldsymbol{G_n}}+\text{h.c.},
\label{Eq:SMHamK}
\end{equation}
where $N_k$ is the number of reciprocal vectors $\boldsymbol{G_n}$ considered and $h_m(\boldsymbol{k})$ is the resulting coupling between states of the two sublattices with momentum difference $\boldsymbol{G_m}$, in terms of the Fourier coefficients it is given by,
\begin{equation}
h_m(\boldsymbol{k})=f_{x,m}c(k_x)e^{i(k_y+G_m)a/2}+f_{y,m}e^{-i(k_y+G_m)a}.
\label{Eq:SMCoupling}
\end{equation}

\begin{figure*}[ht]
\begin{center}
{\includegraphics[width=0.9\textwidth]{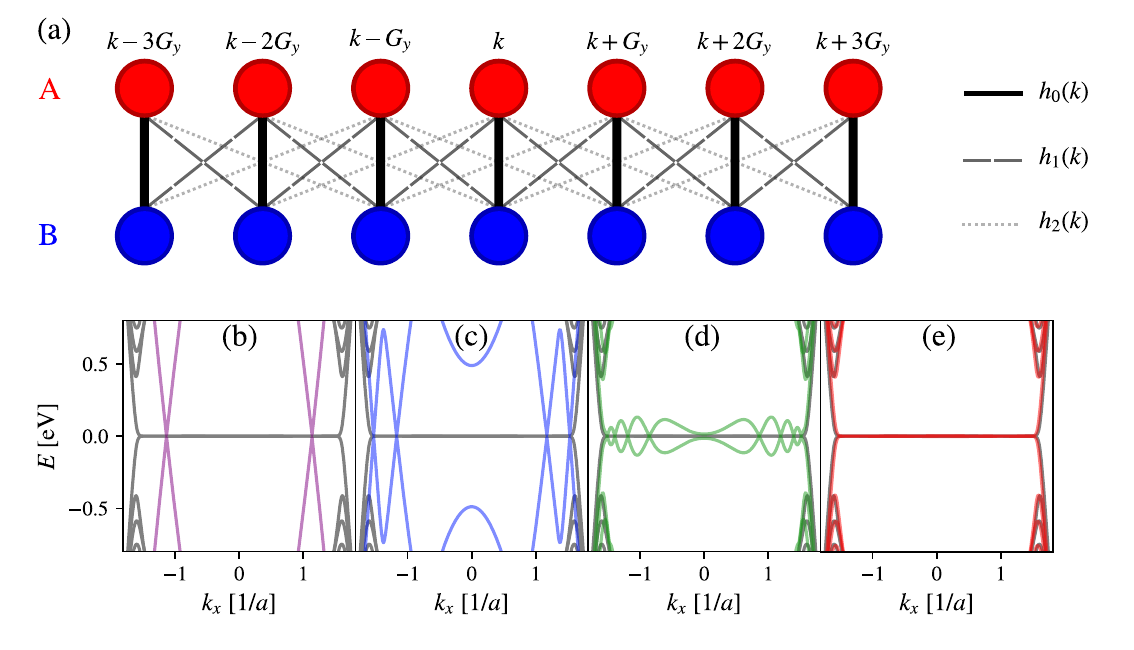}}
\caption{(a) Schematic of the lattice in momentum space with $N_k=7$. It consists of a linear chain with two sublattices indicated by the red and blue circles for sublattice A and B respectively. The lines correspond to the three different types of hoppings given by Eq. (\ref{Eq:SMCoupling}). There are only interlattice hoppings between states separated up to $2G_y$. (b)-(e) Comparison between the spectrum obtained from the momentum space Hamiltonian in Eq. (\ref{Eq:SMHamK}) and that of the real space Hamiltonian (\ref{Eq:HamReal}) for different values of $N_k$. (b) $N_k=1$, (c) $N_k=3$, (d) $N_k=9$ and (e) $N_k=21$. In each case the spectrum of the real space Hamiltonian is shown in gray. The band structures shown have parameters $A=0.15a$ and $\lambda=13.6$nm.
}
\label{fig:SMMomentumLattice}
\end{center}
\end{figure*}

The Hamiltonian in momentum space can be seen as a linear chain with $N_k$ cells and two sublattices, with intersublattice hoppings $h_m$ as shown in Fig.~\ref{fig:SMMomentumLattice}(a). The Hamiltonian can then be diagonalized so that we work within the mBZ with the explicit components of each reciprocal lattice vector,
\begin{equation}
    \mathcal{H}(\boldsymbol{k})\ket{\Phi_{l,\boldsymbol{k}}}=E_{l,\boldsymbol{k}}\ket{\Phi_{l,\boldsymbol{k}}}, \quad
\ket{\Phi_{l,\boldsymbol{k}}}=
\begin{pmatrix}
\vdots\\
\phi^A_{l,\boldsymbol{k}}(\boldsymbol{G_{-1}})\\
\phi^B_{l,\boldsymbol{k}}(\boldsymbol{G_{-1}})\\
\phi^A_{l,\boldsymbol{k}}(\boldsymbol{G_{0}})\\
\phi^B_{l,\boldsymbol{k}}(\boldsymbol{G_{0}})\\
\phi^A_{l,\boldsymbol{k}}(\boldsymbol{G_{+1}})\\
\phi^B_{l,\boldsymbol{k}}(\boldsymbol{G_{+1}})\\
\vdots
\end{pmatrix}.
\end{equation}
In principle $N_k$ should be equal to the number of  atomic sites within the unit cell,  but as we are only interested in the flat bands that appear at low energies, we can truncate the momentum lattice around the Dirac cones until we recover the low energy spectrum. In Fig.~\ref{fig:SMMomentumLattice}(b)-(e) we compare the spectrum obtained for different values of $N_k$ with that of the real lattice. We can see that the flat bands are already captured with $N_k=21$ (42 momentum lattice sites), while the real system has 128 atomic sites.

\section{Hartree Potential}
In the following we derive a Hartree Hamiltonian. This is performed by including the following term in the unperturbed Hamiltonian,
\begin{subequations}
\begin{equation}
    V_H=\int  d^2\boldsymbol{r} v_H(\boldsymbol{r}) \boldsymbol{\Psi^{\dagger}}(\boldsymbol{r}) \boldsymbol{\Psi}(\boldsymbol{r}),
\end{equation}
\begin{equation}
    v_H(\boldsymbol{r})=\int d^2\boldsymbol{r'} v_C(\boldsymbol{r}-\boldsymbol{r'})\delta \rho(\boldsymbol{r'}),
\end{equation}
\end{subequations}
where $\delta \rho(\boldsymbol{r})$ is the fluctuation with respect to the probability density at charge neutrality (CN) $\rho_{CN}(\boldsymbol{r})$ and $v_C(\boldsymbol{r})=\frac{e^2}{\varepsilon|\boldsymbol{r}|}$ is the Coulomb potential with $e$ being the electron charge and $\varepsilon$ the dielectric constant. As the charge distribution has the symmetry of the moir\'e superlattice, an expansion into plane waves is convenient, 
\begin{equation}
    v_H(\boldsymbol{r})=\sum_{\boldsymbol{G}}  v_C(\boldsymbol{G}) \delta \rho(\boldsymbol{G}) e^{i\boldsymbol{G} \cdot \boldsymbol{r}},
\end{equation}
where $\boldsymbol{G}=l_x\boldsymbol{G_x} + l_y\boldsymbol{G_y}$ is a reciprocal lattice vector with $l_x,l_y \in \mathbb{Z}$, $v_C(\boldsymbol{G})$ and $\delta \rho (\boldsymbol{G})$ are the Fourier transform of the Hartree potential and the charge density  respectively, they are given by,
\begin{subequations}
\begin{equation}
    v_C(\boldsymbol{G})=\frac{2\pi e^2}{ \varepsilon A_c |\boldsymbol{G}| },
\end{equation}
\begin{equation}
    \delta \rho(\boldsymbol{G})=2\int_{\text{mBZ}}\frac{d^2k}{V_{\text{mBZ}}}\sum_{i,l,\boldsymbol{G'}}\phi_{l,k}^{i*}(\boldsymbol{G'})\phi_{l,k}^{i}(\boldsymbol{G'}+\boldsymbol{G}),
    \label{Eq:dRhoSM}
\end{equation}
\end{subequations}
here $A_c=\sqrt{3}a\lambda$ is the area of the supercell, $V_{\text{mBZ}}$ is the area of the mBZ and $\phi_{l,\boldsymbol{k}}^i(\boldsymbol{G})$ is the wavefunction component for an electron in a state $l$ of sublattice $i$ with momentum $\boldsymbol{k+G}$. The factor of $2$ is to take into account the spin degeneracy. As we are considering the fluctuation from $\rho_{CN}(\boldsymbol{r})$, only the states with energy $E_l(\boldsymbol{k})$ such that $CN>E_l(\boldsymbol{k}) \geq E_f $ are taken in Eq.~(\ref{Eq:dRhoSM}).

This can be further simplified by considering that the superlattice wavelengths along the $y$ direction are much greater than those in the $x$ direction. This is reflected in the magnitude of the Fourier coefficients of the Coulomb potential,
\begin{equation}
    v_C(\boldsymbol{G_x})=\frac{e^2}{\lambda\varepsilon}, \quad 
    v_C(\boldsymbol{G_y})=\frac{e^2}{\sqrt{3}a \varepsilon }, 
\end{equation}
as $v_C(\boldsymbol{G_x})/v_C(\boldsymbol{G_y})=\sqrt{3}a/\lambda<<1$, only the Fourier components along the $y$-direction are relevant, manifesting the quasi-one dimensionality of the system.  The potential can then be redefined as,
\begin{subequations}
\begin{equation}
    v_H(y)=\sum_n v_C(G_n)\delta\rho(G_n)e^{iG_ny},
\end{equation}
\begin{equation}
    v_C(G_n)=\frac{e^2}{\varepsilon\sqrt{3}a|n|},
\end{equation}

\begin{equation}
    \delta \rho(G_n)=2\int_{\text{mBZ}}\frac{d^2k}{V_{\text{mBZ}}}\sum_{i,l,m}\phi_{l,k}^{i*}(G_m)\phi_{l,k}^{i}(G_m+G_n).
    \label{Eq:derhonSM}
\end{equation}
\end{subequations}
By adding the contribution of the Hartree term, the Hamiltonian in momentum space becomes,
\begin{equation}
    \mathcal{H'}=\mathcal{H}+\sum_{\boldsymbol{k}} \sum_{n}^{N_k} v_C(G_n)  \delta \rho(G_n) \bigl[a_{\boldsymbol{k}+\boldsymbol{G_n}}^{\dagger} a_{\boldsymbol{k}}
    +b_{\boldsymbol{k}+\boldsymbol{G_n}}^{\dagger}b_{\boldsymbol{k}}\bigr].
    \label{Eq:Ham+HarSM}
\end{equation}
Now $\delta \rho (G_n)$ is calculated in a self-consistent scheme. We consider a fixed filling $\nu$ with respect to CN such that $E_F$ is recalculated within each iteration. We start by diagonalizing the non-interacting Hamiltonian in Eq. (\ref{Eq:SMHamK}) to obtain $\phi_{l,k}^{i}(G_n)$, CN and $E_f$ for the desired filling $\nu$. These values are used to calculate the first iteration of $\delta \rho (G_n)$ according to the Eq. (\ref{Eq:derhonSM}), which enters the interacting Hamiltonian of Eq. (\ref{Eq:Ham+HarSM}), then by diagonalizing again new values for $\phi_{l,k}^{i}(G_n)$, CN and $E_F$ are obtained, repeating the process until convergence is achieved. We note that by considering different seeds other metastable solutions are found. 

\section{Hartree Potential Strength}

\begin{figure*}[ht]
\begin{center}
{\includegraphics[width=0.9\textwidth]{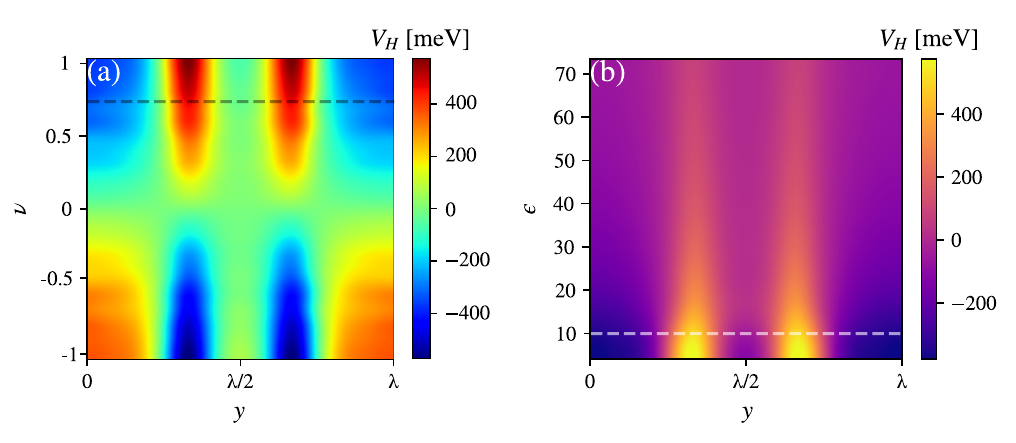}}
\caption{Map of the Hartree potential as a function of $y$ and (a) filling fraction (b) dielectric constant for a system with $A=0.15a$ and $\lambda=13.6$ nm. The black dashed line in (a) indicates $\nu=0.75$, the value used for (b), while the white dashed line in (b) $\varepsilon=10$ is the value used on (a).}
\label{fig:HartreeMaps}
\end{center}
\end{figure*}

In Fig.~\ref{fig:HartreeMaps} we show the strength of the electrostatic Hartree potential. In Fig.~\ref{fig:HartreeMaps}(a), as a function of filling, the Hartree potential reaches values as high as $500$ meV with narrow bands of a few meV. In Fig. ~\ref{fig:HartreeMaps}(b) we show the effect of the dielectric environment.  

\section{Electrostatic broken-symmetry states}

In this section we quantify the energetic status of the one-cell and enlarged-cell Hartree solutions discussed in the main text. This comparison is needed to distinguish the lowest-energy self-consistent state from higher-energy self-consistent textures obtained from enlarged-cell initial conditions. The total mean-field energy is calculated as~\cite{Cea2020band},

\begin{equation}
    E=\sum_{\boldsymbol{k},l}E_{l}(\boldsymbol{k})-\frac{1}{2} \braket{H_H},
\end{equation}
where the sum runs over the occupied states at fixed filling, and the second term removes the double counting of the Hartree contribution already included in the self-consistent single-particle eigenvalues.

We use the term metastable in an operational self-consistent mean-field sense. A metastable state is a converged Hartree fixed point reached from an enlarged-cell seed that remains stable under iteration, but has a higher total energy  than the lowest-energy converged solution at the same parameters and filling. Thus, metastability here does not imply that the enlarged-cell texture is the ground state. It means that the texture is a stable self-consistent Hartree solution. We first compare the one-cell SS and sublattice-SP solutions for $A=0.15a$, $\lambda=13.6$ nm, and $\varepsilon=10$. The corresponding energies are shown in Table~\ref{tab:1CellEn}. For these parameters, the SS solution has the lowest energy over the filling range shown.

\begin{table}[htbp]
\setlength\extrarowheight{5pt}
\begin{center}
\begin{tabular}{ |c|c|c|c|c|c|c|c|c|c|c| } 
\hline
$\nu$ & 0.1 & 0.2 & 0.3 & 0.4 & 0.5 & 0.6 & 0.7 & 0.8 & 0.9 & 1\\
\hline
SS Energy [meV]& 1.12 & 6.05 & 15.47 & 22.21 & 36.32 & 48.76 & 59.09 & 77.30 & 91.36 & 111.42\\
\hline
SP Energy [meV] & 2.01 & 7.77 & 15.92 & 29.35 & 43.89 & 59.07 & 75.45 & 94.38 & 112.84 & 134.50\\
\hline
\end{tabular}
\end{center}
\caption{Total mean-field energy, in meV, for the one-cell SS and SP solutions as a function of filling.}
\label{tab:1CellEn}
\end{table}
we now extend the Hartree calculation to enlarged supercells with two and three moir\'e unit cells. This allows us to test whether reduced-periodicity charge textures can become energetically favored. In the two-cell case, we consider $\nu=1/2$, corresponding to one extra electron distributed over two moir\'e cells. We initialize the Hartree iteration with density seeds containing a Fourier component at the reduced reciprocal vector $\boldsymbol{G}_{1/2}$, with different phases. These seeds allow the iteration to explore solutions that are not constrained to repeat after one moir\'e cell.

\begin{figure}[ht]
\begin{center}
{\includegraphics[scale = 0.9]{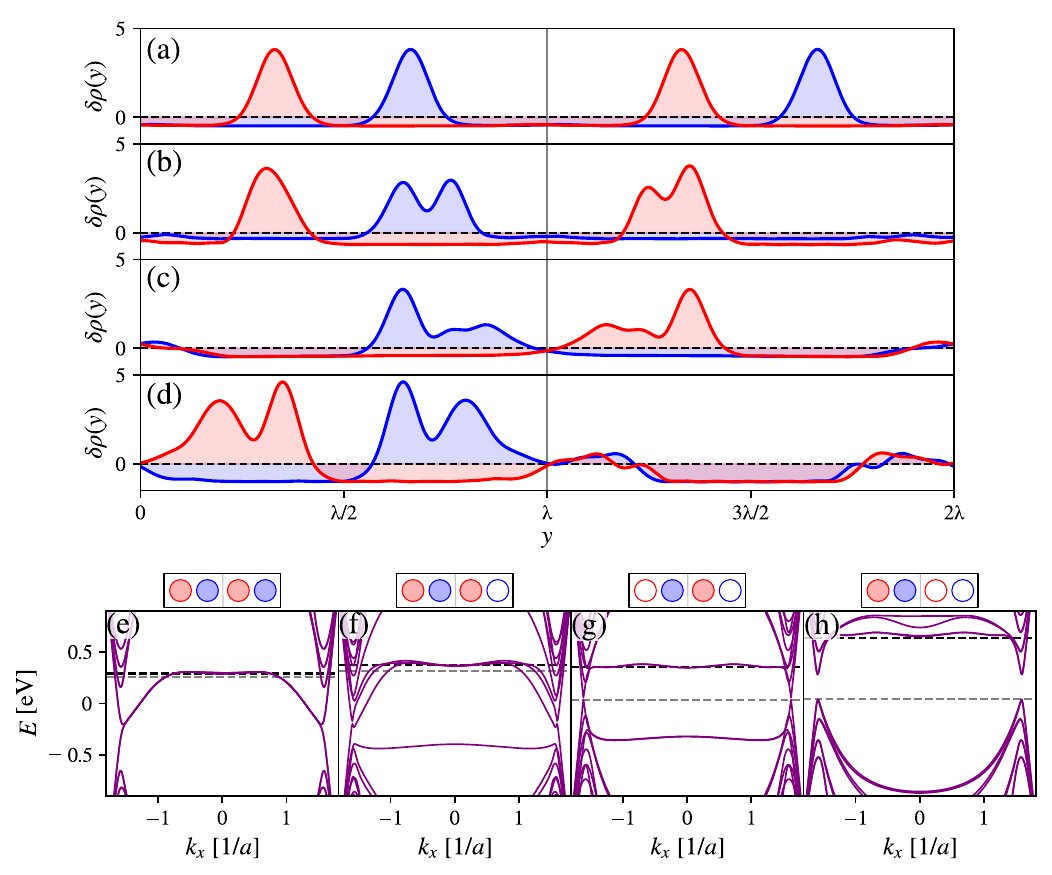}}
\caption{Charge density of the self-consistent solutions found considering two unit cells for $\nu=1/2$ (one extra electron between the two cells) with $A=0.15a$, $\lambda=13.6$ nm and $\varepsilon=10$. (a) One-cell periodic solution represented in a two-cell supercell. (b)-(d) Solutions with reduced translational symmetry. The solutions are ordered in ascending energy. Panel (a) is the lowest-energy solution among the converged states compared here, while panels (b)--(d) are higher-energy self-consistent solutions that remain stable under iteration.} (e)-(h) The corresponding band structure for (a)-(d), the gray and the black dashed lines indicate the charge neutrality and the Fermi energy for each case. An inset of a schematic of the localization is shown for each band structure.
\label{fig:2Cells}
\end{center}
\end{figure}

Figure~\ref{fig:2Cells} shows representative converged two-cell solutions. Panel (a) is the one-cell periodic solution represented in a doubled supercell. Panels (b)--(d) are reduced-periodicity self-consistent textures. In panel (b), one cell is polarized predominantly on one sublattice. In panel (c), the two cells display opposite sublattice polarization. In panel (d), most of the additional charge is concentrated in one of the two cells.

The corresponding energies are shown in Table~\ref{tab:2CellEn}. The energy difference $\Delta E$ is measured relative to the one-cell periodic solution shown in Fig.~\ref{fig:2Cells}(a). Among the converged two-cell solutions compared here, the one-cell periodic solution has the lowest energy. The reduced-periodicity textures are therefore metastable in the sense defined above.

\begin{table}[htbp]
\setlength\extrarowheight{5pt}
\begin{center}
\begin{tabular}{ |c|c|c|c| } 
\hline
Solution  & Seed $\delta \rho(\boldsymbol{G}_{1/2})$ & Energy per cell [meV] & $\Delta E$ [meV]\\
\hline
(a) One-cell periodic & 0 & 37.21 & 0 \\ 
\hline
(b) Two-cell periodic & $0.1 e^{i\frac{5\pi}{4}}$ & 49.88 & 11.67\\ 
\hline
(c) Two-cell periodic & $0.1e^{i\pi}$ & 73.73 & 35.52 \\ 
\hline
(d) Two-cell periodic & $0.1 e^{i\frac{3\pi}{2}}$ & 107.79 & 69.58 \\ 
\hline
\end{tabular}
\end{center}
\caption{Energies of the self-consistent solutions found in a two-cell calculation at $\nu=1/2$. The labels correspond to those used in Fig.~\ref{fig:2Cells}. The energy difference $\Delta E$ is measured relative to the one-cell periodic solution represented in the enlarged supercell.}
\label{tab:2CellEn}
\end{table}

\begin{figure}[ht]
\begin{center}
{\includegraphics[scale = 0.9]{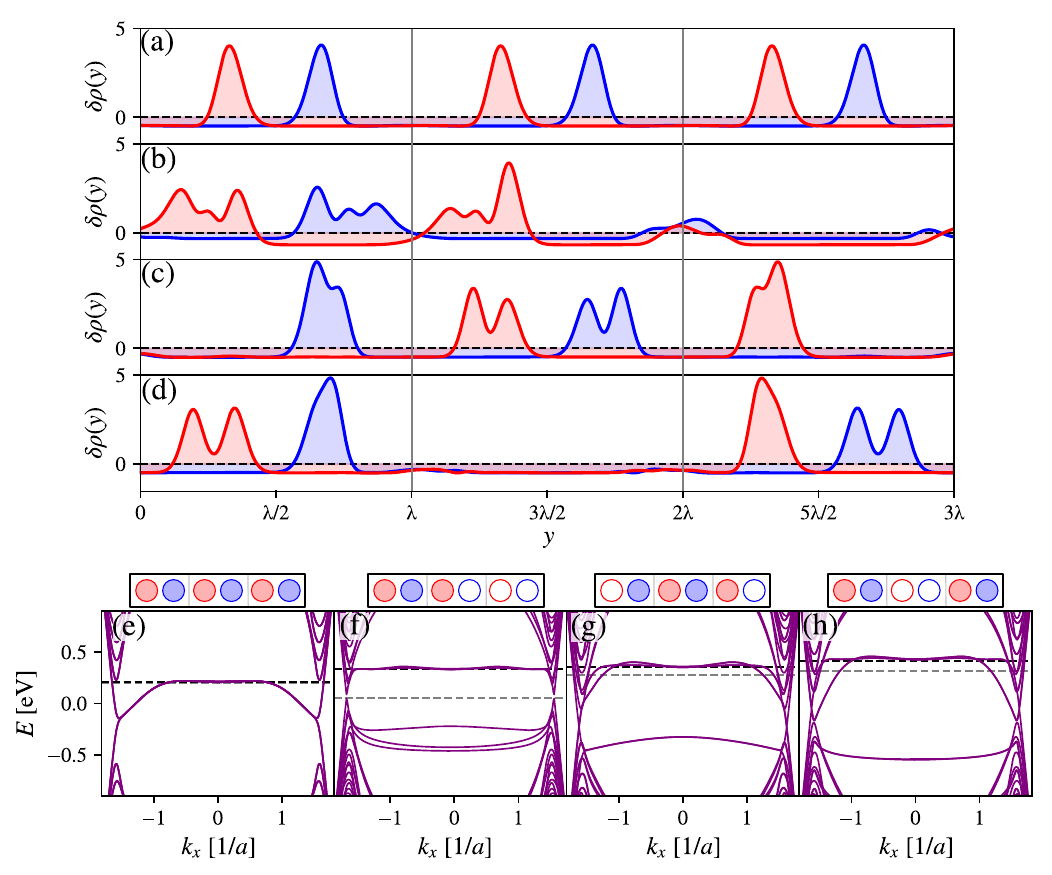}}
\caption{Charge density of the self-consistent solutions found considering three unit cells for $\nu=1/3$ (one extra electron between the three cells) with $A=0.15a$, $\lambda=13.6$nm and $\varepsilon=10$. (a) One-cell periodic solution represented in a three-cell supercell. (b)-(d) Solutions with reduced translational symmetry. The solutions are ordered in ascending energy. Panel (a) is the lowest-energy solution among the converged states compared here, while panels (b)--(d) are higher-energy self-consistent solutions that remain stable under iteration. (e)-(h) The corresponding band structure for (a)-(d), the gray and the black dashed lines indicate the charge neutrality and the Fermi energy for each case. An inset of a schematic of the localization is shown for each band structure.}
\label{fig:3Cells}
\end{center}
\end{figure}

We perform the same analysis for a three-cell supercell at $\nu=1/3$, corresponding to one extra electron distributed over three moir\'e cells. The Hartree iteration is initialized with density seeds containing a Fourier component at the reduced reciprocal vector $\boldsymbol{G}_{1/3}$. Representative converged solutions are shown in Fig.~\ref{fig:3Cells}. As in the two-cell case, panel (a) is the one-cell periodic solution represented in an enlarged supercell, while panels (b)--(d) are reduced-periodicity textures.

Table~\ref{tab:3CellEn} gives the corresponding energies. Again, the energy difference $\Delta E$ is measured relative to the one-cell periodic solution. Among the converged three-cell solutions compared here, the lowest-energy state preserves the one-cell translational periodicity. The three-cell textures are therefore higher-energy metastable Hartree fixed points.

\begin{table}[htbp]
\setlength\extrarowheight{5pt}
\begin{center}
\begin{tabular}{ |c|c|c|c|c| } 
\hline
Solution  & Seed $\delta \rho(\boldsymbol{G}_{1/3})$ & Energy per cell [meV] & $\Delta E$ [meV]\\
\hline
(a) One-cell periodic & 0 & 16.92 & 0 \\ 
\hline
(b) Three-cell periodic & $0.1 e^{i\frac{5\pi}{4}}$ & 30.38 & 13.46\\ 
\hline
(c) Three-cell periodic & $0.1e^{i\pi}$ & 40.99 & 24.07 \\ 
\hline
(d) Three-cell periodic & $0.1$ & 51.71 & 34.79 \\ 
\hline
\end{tabular}
\end{center}
\caption{Energies of the self-consistent solutions found in a three-cell calculation at $\nu=1/3$. The labels correspond to those used in Fig.~\ref{fig:3Cells}. The energy difference $\Delta E$ is measured relative to the one-cell periodic solution represented in the enlarged supercell.}
\label{tab:3CellEn}
\end{table}

For both the two-cell and three-cell calculations, we tested several Hartree seeds with different amplitudes and phases at the reduced reciprocal vectors $\boldsymbol{G}_{1/2}$ and $\boldsymbol{G}_{1/3}$. The tables show representative converged solutions. Other seeds either returned to the one-cell periodic state or produced similar reduced-periodicity textures with higher energy. Thus, within the set of converged states compared here, the lowest-energy solution retains the translational symmetry of one moir\'e unit cell. This energetic comparison fixes the role of the enlarged-cell textures in the manuscript. They are not used as parent states for the superconducting calculation. The pairing analysis in the main text is performed using the one-cell SS and SP reconstructed normal states, which are the relevant lowest-energy parent states for the parameters considered here. The enlarged-cell solutions instead demonstrate that the Hartree landscape contains locally stable electrostatic textures with reduced translational symmetry.

\section{Screening of the Coulomb Potential}
To calculate the screened Coulomb potential in momentum space, we first consider a double-gate screened Coulomb potential,
\begin{equation}
    v^{dg}_C(\boldsymbol{q})=2 \pi e^2 \frac{\text{tanh}(d_g |\boldsymbol{q}|)}{A_c \varepsilon |\boldsymbol{q}|},
    \label{eq:vc(q)}
 \end{equation}
where $d_g=40$ nm is the distance between the sample and a metallic gate. We will proceed to calculate the electronic susceptibility within the random phase approximation (RPA). As we will see in the following, working within the mBZ of a moir\'e system is more complex than a non-moir\'e one due to the presence of Umklapp processes. First, consider two states, one in band $l_1$ with momentum $\boldsymbol{k}$, and the other in band $l_2$  with momentum $\boldsymbol{k+q}$. We define $\Upsilon_{\boldsymbol{k},\boldsymbol{k+q}}^{l_1,l_2}(\boldsymbol{G_n})$ as the sum of the overlaps between components separated by $\boldsymbol{G_n}$ between these states, i.e.,
\begin{equation}
\Upsilon_{\boldsymbol{k},\boldsymbol{k+q}}^{l_1,l_2}(\boldsymbol{G_n})=\sum_{i,m} \phi^{i*}_{l_1,\boldsymbol{k}}(\boldsymbol{G_{m}})\phi^{i}_{l_2,\boldsymbol{k+q}}(\boldsymbol{G_m+G_n}), 
\end{equation}
the electronic susceptibility can then be written as,
\begin{equation}
    \hat{\chi}(\boldsymbol{q})_{m,n}= \frac{2}{N} \sum_{\boldsymbol{k},l_1,l_2} \frac{f(E_{l_1,\boldsymbol{k}})-f(E_{l_2,\boldsymbol{k+q}})}{E_{l_1,\boldsymbol{k}}-E_{l_2,\boldsymbol{k+q}}}
    \Upsilon_{\boldsymbol{k},\boldsymbol{k+q}}^{l_1,l_2}(\boldsymbol{G_m})
    \Upsilon_{\boldsymbol{k},\boldsymbol{k+q}}^{l_1,l_2*}(\boldsymbol{G_n})
    ,
\end{equation}
where $N$ is the number of $k$-points and a factor of 2 is included to account for spin degeneracy. Here $f(E)=(1+e^{(E-E_F)/k_BT})^{-1}$ is the Fermi-Dirac distribution with $T$ being the temperature and $k_B$ Boltzmann constant. The electronic susceptibility is a matrix whose components $\hat{\chi}(\boldsymbol{q})_{m,n}$ correspond to different Umklapp processes between states $\boldsymbol{k+G_m}$ and $\boldsymbol{k+G_n+q}$. The screening of the Coulomb potential due to particle-hole excitations is then computed within the RPA limit,
\begin{equation}
    \hat{V}_{\text{scr}}(\boldsymbol{q})=\left(\hat{V_C}^{-1}(\boldsymbol{q})-\hat{\chi}(\boldsymbol{q})\right)^{-1},
    \label{Eq:ScreenedV}
\end{equation}
where $\hat{V}_C(\boldsymbol{q})$ is a diagonal matrix whose elements are given by,
\begin{equation}
    \hat{V}_{C}(\boldsymbol{q})_{m,n}=v^{dg}_C(\boldsymbol{q}+\boldsymbol{G_m}) \delta_{m,n}.
    \label{Eq:Coulomb}
\end{equation}
\begin{figure}[ht]
\begin{center}
{\includegraphics[scale = 0.92]{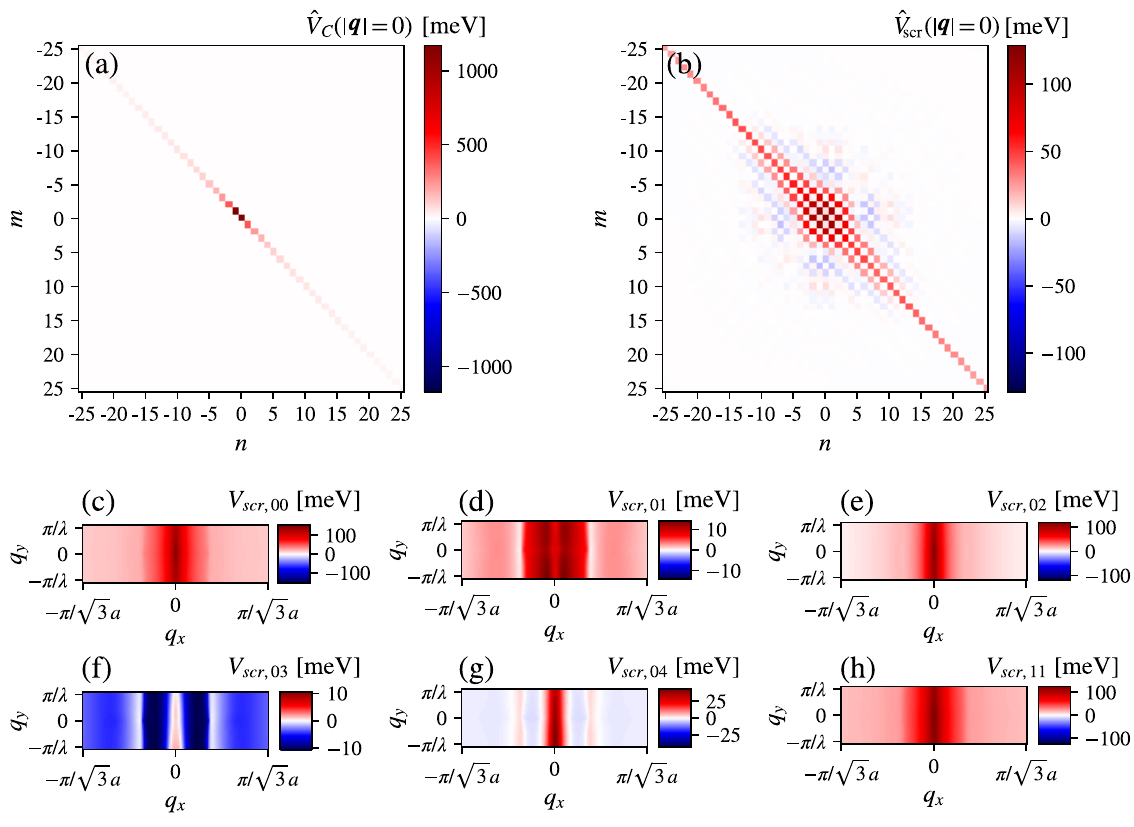}}
\caption{Screening of the Coulomb potential in reciprocal space for the NH case with $A=0.15a$, $\lambda=13.6$ nm and $\varepsilon=10$.  (a) Matrix elements of the Coulomb potential matrix without screening Eq. (\ref{Eq:Coulomb}). (b) Matrix elements of the screened Coulomb potential Eq. (\ref{Eq:ScreenedV}). In both cases $|\boldsymbol{q}|=0$ was considered. (c)-(h) Screened Coulomb potential as a function of $\boldsymbol{q}$ for different Umklapps channels. 
}
\label{fig:Screening}
\end{center}
\end{figure}

In Fig.~\ref{fig:Screening}(a)-(b) we show the matrix elements of the Coulomb potential matrix before and after the screening, respectively. One thing to notice is the diminishing magnitude, going from a maximum value of $\sim1.1$eV to $\sim120$meV after screening. Furthermore, the screening opens up non-diagonal channels, some of which are negative. Fig.~\ref{fig:Screening}(c)-(h) show the dependence on $\boldsymbol{q}$ for different Umklapp processes of the screened potential. 

\section{Kohn--Luttinger RPA Superconductivity}
Now that we have the screened potential in momentum space, we determine the critical scale of the linearized pairing instability ($T_C$) within a framework similar to that of Kohn--Luttinger~\cite{Cea2021Coulomb,Long2024Evolution}. We start with the linearized gap equation for the order parameter (OP),
\begin{equation}
    \Delta_{l_1,l_2}(\boldsymbol{k})=\sum_{\boldsymbol{k'}}\sum_{l_1',l_2'}\Gamma_{l_1,l_2}^{l_1',l_2'}(\boldsymbol{k,k'}) \Delta_{l_1',l_2'}(\boldsymbol{k'}),
\end{equation}
where $\Gamma_{l_1,l_2}^{l_1',l_2'}$ is the superconducting kernel, it is given by,
\begin{equation}
    \Gamma_{l_1,l_2}^{l_1',l_2'}(\boldsymbol{k},\boldsymbol{k'})=-\frac{1}{N}
    \mathcal{F}(E_{l_1,\boldsymbol{k}},E_{l_2,\boldsymbol{k}})
    \mathcal{F}(E_{l_1',\boldsymbol{k'}},E_{l_2',\boldsymbol{k'}}) 
    \sum_{m,n}
    \hat{V}_{\text{scr}}(\boldsymbol{k'-k},\boldsymbol{G_m},\boldsymbol{G_n})
\Upsilon_{\boldsymbol{k},\boldsymbol{k'}}^{l_1,l_1'*}(\boldsymbol{G_m})
\Upsilon_{\boldsymbol{k},\boldsymbol{k'}}^{l_2,l_2'}(\boldsymbol{G_n}),
\end{equation}
where
\begin{equation}
     \mathcal{F}(E_1,E_2)=\sqrt{\frac{1-f(E_2-E_F)-f(E_1-E_F)}{E_1+E_2-2E_F}},
\end{equation}
\begin{figure}[ht]
\begin{center}
{\includegraphics[scale = 0.95]{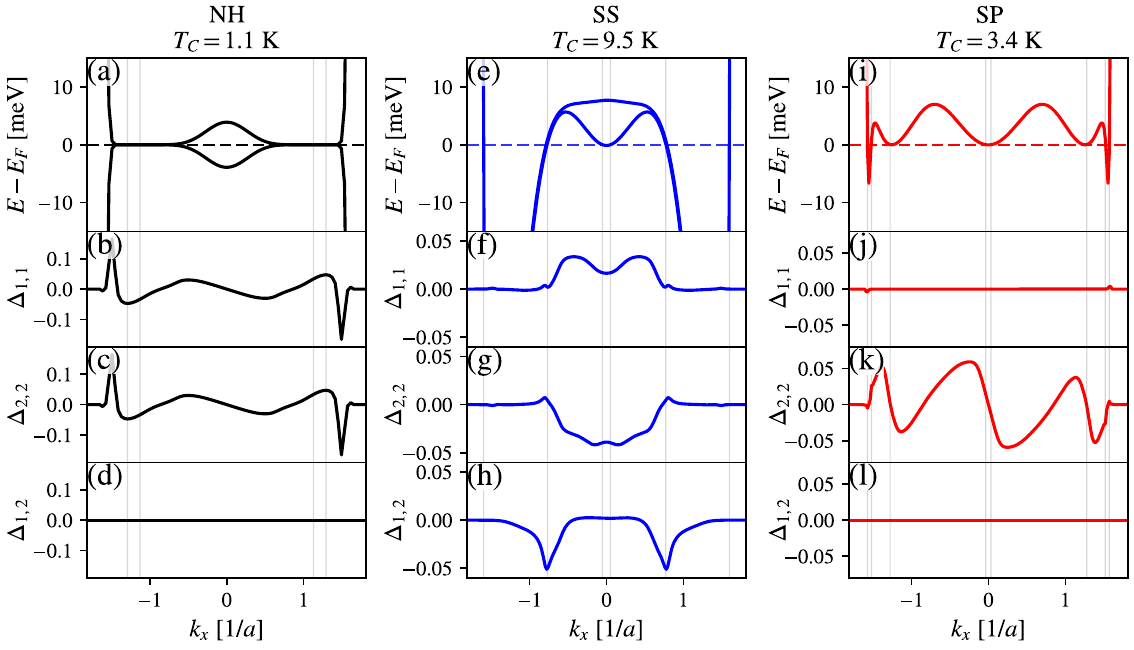}}
\caption{Band structure and OPs at $\nu=0.1$ with $A=0.15a$, $\lambda=13.6$ nm and $\varepsilon=10$ at their respective $T_C$, for (a)-(d) NH, (e)-(h) SS and (i)-(l) SP. For each case we show the intraband projections for each band $\Delta_{1,1}$, $\Delta_{2,2}$ and the interband projection $\Delta_{1,2}$. The dashed horizontal lines indicate the Fermi level, while the solid vertical ones indicate the Fermi surface.}
\label{fig:OP}
\end{center}
\end{figure}
where $\hat{V}_{\text{scr}}(\boldsymbol{q},\boldsymbol{G_m},\boldsymbol{G_n})=\hat{V}_{\text{scr}}(\boldsymbol{q})_{m,n}$. The kernel is projected onto the two bands at the middle of the spectrum, these bands will be labeled as band 1 and band 2, in ascending energy order. The kernel is then diagonalized. The critical scale $T_C$ is defined by the temperature (which enters implicitly through the Fermi-Dirac distribution) at which the largest eigenvalue of the kernel is equal to 1. red Notice that the Hartree calculation is used to reconstruct the normal state band structure and the superconducting calculation is then performed on top of this reconstructed normal state. The interaction entering the Kohn--Luttinger kernel is the screened interaction in the particle-hole channel, evaluated within RPA and projected into the Cooper channel. It is not added again as a static Hartree correction to the one-particle Hamiltonian. Thus, the Hartree term is used only to define the reconstructed normal-state propagators. The anomalous Cooper-channel instability is then evaluated on top of this fixed normal state. We do not add a second static Hartree self-energy in the gap equation, and the superconducting order parameter is not fed back into the Hartree self-consistency loop.

We calculated the resulting $T_C$ as a function of $\nu$ for the NH, SS and SP cases, as shown in Fig.~4 in the main text. The largest value appears for the SS case, reaching $9.5$ K at $\nu=0.1$. In Fig.~\ref{fig:OP} we show the band structures and OPs for the three cases considered at $\nu=0.1$. For each case, the OPs are shown at the corresponding $T_C$. In the NH case, since for this value of $\nu$ the $E_F$ is almost equal to the CN point and the spectrum is symmetric around it, the OP looks the same on each band. The OP features peaks around $k_xa\approx\pm1.6$, corresponding to the points where the bands cease to be flat. For the SS case, the OP is distributed between both bands but there is a sign change between them. Additionally, we have a non-zero interband OP at the Fermi surface where the bands are degenerate. Finally, in the SP case, the OP is mostly concentrated on band 2 since the Fermi surface lies solely within this band and is separated by a gap from band 1. Within the spin-compensated parent states considered here, the antisymmetric OPs in the NH and SP cases correspond to spin-triplet pairing, while the symmetric OP in the SS case corresponds to spin-singlet pairing.

\section{Quantum Geometry}

Quantum geometry can affect superconducting properties in flat-band systems, in particular through its contribution to the superfluid weight~\cite{torma2022superconductivity,yu2025quantum}. In the present work we use the quantum metric only as a diagnostic of the momentum-space structure of the flat-band wave functions. We do not use it to compute the transition temperature.

\begin{figure}[hbtp!]
\begin{center}
{\includegraphics[scale = 0.92]{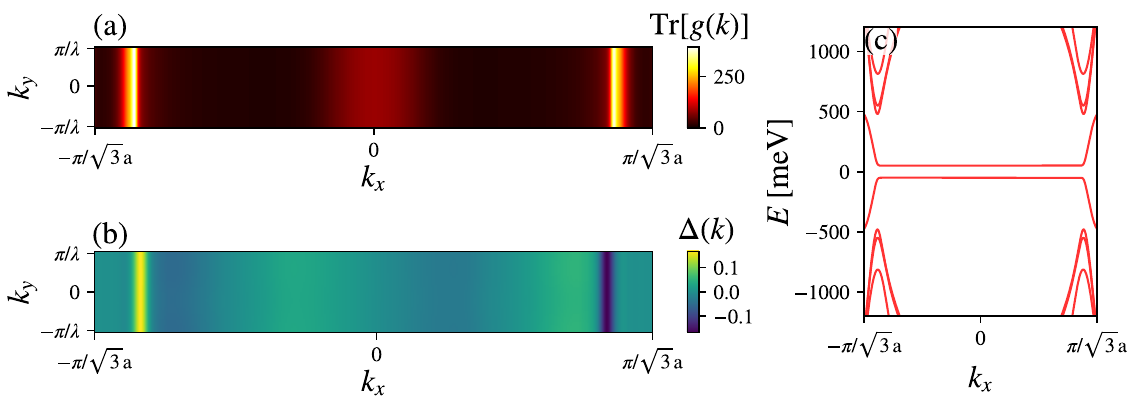}}
\caption{ Interplay between the quantum geometry and superconductivity. (a) Trace of the regularized quantum metric computed with $\delta_{SL}=5$ meV and $\delta_{SW}=4$ meV. (b) OP for the NH case at $\nu=0.1$ as shown in Fig.~\ref{fig:OP}. (c) Isolated flat bands through the inclusion of a mass term and a short-wavelength modulation. Panels (a) and (b) use the weak auxiliary perturbations described in the text to regularize the metric. Panel (c) uses larger values only to make the band isolation visible in the plot ($\delta_{SL}=50$ meV and $\delta_{SW}=6$ meV).}
\label{fig:QuantumMetric}
\end{center}
\end{figure}

For an isolated band, the quantum geometric tensor is
\begin{equation}
    Q_{ij}(\boldsymbol{k})=
    \bra{\partial_{k_i}u_{\boldsymbol{k}}}
    \left[\mathds{1}-\ket{u_{\boldsymbol{k}}}\bra{u_{\boldsymbol{k}}}\right]
    \ket{\partial_{k_j}u_{\boldsymbol{k}}},
\end{equation}
and its symmetric part defines the quantum metric,
\begin{equation}
    g_{ij}(\boldsymbol{k})=\frac{Q_{ij}(\boldsymbol{k})+Q_{ji}(\boldsymbol{k})}{2}.
\end{equation}
In the unperturbed strained-monolayer spectrum, the flat bands are not globally isolated because they approach the dispersive bands near the mini-Brillouin-zone boundary. A direct single-band metric is therefore not a well-defined global quantity. To obtain a regularized diagnostic, we compute a regularized metric diagnostic for the isolated flat-band manifold by adding two auxiliary perturbations,
\begin{equation}
    \delta H_{SL}=\sum_{n}^{N_k} \delta_{SL}
    \left[
    a_{\boldsymbol{k}+\boldsymbol{G_n}}^{\dagger}a_{\boldsymbol{k}+\boldsymbol{G_n}}
    -
    b_{\boldsymbol{k}+\boldsymbol{G_n}}^{\dagger}b_{\boldsymbol{k}+\boldsymbol{G_n}}
    \right],
\end{equation}
and
\begin{equation}
    \delta H_{SW}=\sum_{n,m}^{N_k} \delta_{SW}
    \left[
    a_{\boldsymbol{k}+\boldsymbol{G_n}+\boldsymbol{G_m}}^{\dagger}
    b_{\boldsymbol{k}+\boldsymbol{G_n}}
    +\mathrm{h.c.}
    \right].
\end{equation}
Here $\delta_{SL}=5$ meV and $\delta_{SW}=4$ meV for the metric shown in Fig.~\ref{fig:QuantumMetric}(a). These perturbations are used only to regularize the quantum-metric diagnostic shown in Fig.~\ref{fig:QuantumMetric}; they are not included in the Hartree reconstruction or in the superconducting gap-equation calculation.

Figure~\ref{fig:QuantumMetric}(a) shows the trace of this regularized metric. The largest values appear near the momentum regions where the flat-band wave functions change most rapidly and where the flat bands approach the dispersive bands. Comparing this diagnostic with the NH order parameter in Fig.~\ref{fig:QuantumMetric}(b), we find that the largest pairing amplitude coincides with the same momentum regions. We therefore interpret the metric comparison as a qualitative correlation between the structure of the flat-band wave functions and the momentum dependence of the pairing eigenfunction, rather than as an independent calculation of the superconducting transition temperature.

\section{Interband Contributions to the Superconductivity}

\begin{figure}[ht]
\begin{center}
{\includegraphics[scale = 0.95]{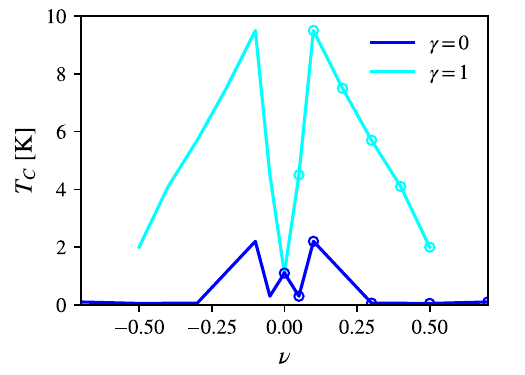}}
\caption{$T_C$ as a function of $\nu$ for the SS case with $\gamma=0$ and $\gamma=1$. In both cases $A=0.15a$, $\lambda=13.6$nm and $\varepsilon=10$ were used.
}
\label{fig:TC_SS}
\end{center}
\end{figure}

In Fig.~\ref{fig:OP} the only case where there is an interband OP is that of the SS case. Interestingly, it is the case with the highest $T_C$. We study the relevance of the interband terms by introducing a parameter $\gamma$ that modulates the terms $l_1 \neq l_2$ and $l_1' \neq l_2'$, such that for $\gamma=0$ the interband terms are turned off and for $\gamma=1$ we return to the original kernel. In Fig.~\ref{fig:TC_SS} we compare the values of the $T_C$ for $\gamma=0$ and $\gamma=1$. It can be seen that without the interband terms the $T_C$ is drastically lower, having a maximum value of $T_C=2.2$ K at $\nu=0.1$, and approaching zero afterwards.

\begin{figure}[hbtp!]
\begin{center}
{\includegraphics[scale = 0.8]{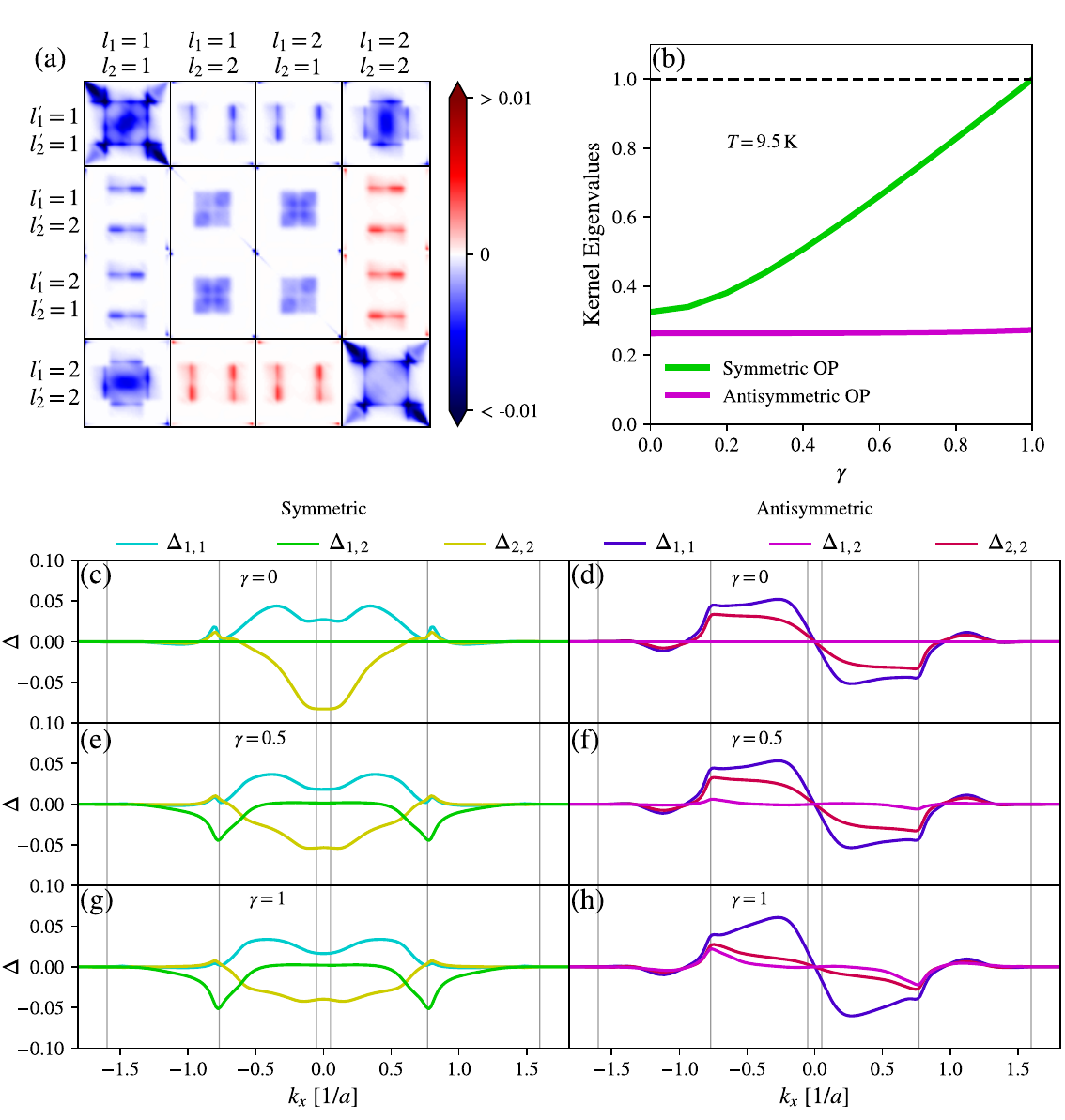}}
\caption{(a) Map of the kernel matrix values for the SS Hartree at $\nu=0.1$ and $T=9.5$ K, here the blocks for the different band processes are separated by the black lines. (b) The two highest eigenvalues of the kernel as a function of the parameter $\gamma$ which modulates the interband processes. In green (purple) we show the eigenvalue with a symmetric (antisymmetric) pairing. (c)-(h) Order parameter projected for intraband processes in each of the two bands considered (band 1 and band 2) and interband processes between them. (c)-(d) For the symmetric and antisymmetric pairing with $\gamma=0$. Similarly (e)-(f) for $\gamma=0.5$ and (g)-(h) for $\gamma=1$. The gray vertical lines indicate where the Fermi surface lies. Here $A=0.15a$, $\lambda=13.6$nm and $\varepsilon=10$ were used.}
\label{fig:OP_Interband}
\end{center}
\end{figure}

To further analyze the importance of the interband processes in Fig.~\ref{fig:OP_Interband}(b) we analyze the evolution of the kernel eigenvalues as a function of $\gamma$. We show the two highest eigenvalues. The first corresponds to the symmetric OP previously shown in Fig.~\ref{fig:OP}. The second is antisymmetric with respect to $k_x$. As $\gamma$ increases, it can be seen that only the eigenvalue corresponding to the symmetric OP increases, while the antisymmetric remains mostly constant. Thus, interband terms greatly increase the difference between the $T_C$ of the symmetric and antisymmetric OPs. In Fig.~\ref{fig:OP_Interband}(c)-(h) we show the evolution of both OPs as a function of $\gamma$. In the symmetric case $\Delta_{1,2}$ increases, becoming the greatest contribution near the Fermi surface. While for the antisymmetric case, the OPs shapes are just slightly different and $\Delta_{1,2}$ always has the smallest contribution.

\section{Hartree-Fock: effects of the Exchange Term}

\begin{figure}[ht]
\begin{center}
{\includegraphics[scale = 0.9]{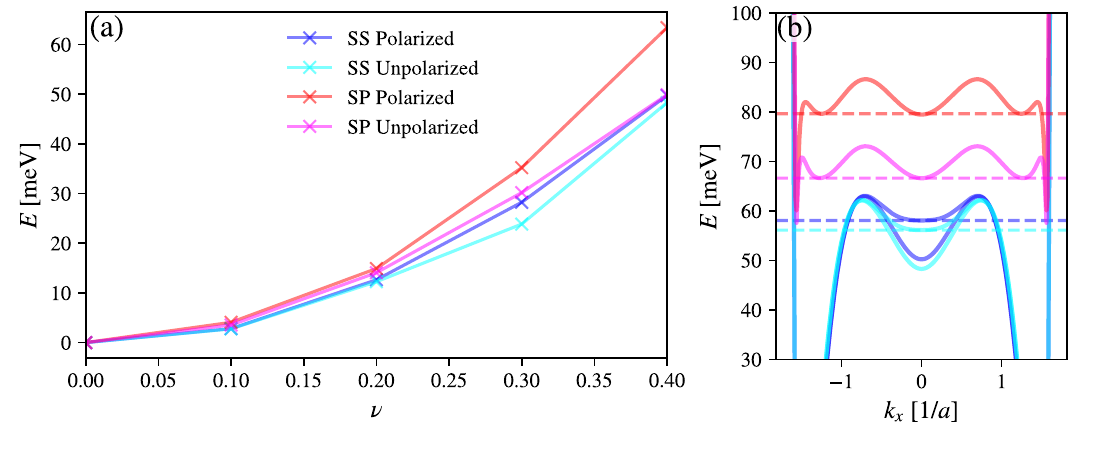}}
\caption{
(a) Total energy as function of $\nu$ for both SS and SP solutions considering both spin polarized and non-polarized solutions. (b) Band structure for each case at $\nu=0.1$.}
\label{fig:HF_Energy}
\end{center}
\end{figure}

Following Ref.~\cite{Cea2020band}, the Fock Hamiltonian can be treated within a plane-wave expansion as was previously done for the Hartree term. The contribution of the exchange term is~\cite{Cea2020band}, 
\begin{subequations}
\begin{equation}
    V_F=\int  d^2\boldsymbol{r} d^2\boldsymbol{r'} \boldsymbol{\Psi^{\dagger}} (\boldsymbol{r})v_F(\boldsymbol{r},\boldsymbol{r'})\boldsymbol{\Psi}(\boldsymbol{r'}),
\end{equation}
For the system studied in present work, it acquires the following form,

\end{subequations}
\begin{equation}
    \mathcal{H}_F=\sum_{\boldsymbol{k}} \sum_{m,n}^{N_k} F_{\sigma,\sigma '}^{i,i'}(\boldsymbol{k},\boldsymbol{G_m},\boldsymbol{G_n}) c_{\boldsymbol{k} + \boldsymbol{G_m}}^{i,\sigma,\dagger} c_{\boldsymbol{k}+\boldsymbol{G_n}}^{i',\sigma'}+\text{h.c.},
\end{equation}
where,
\begin{equation}
    F_{\sigma,\sigma '}^{i,i'}(\boldsymbol{k},\boldsymbol{G_m},\boldsymbol{G_n}) =-\delta_{\sigma,\sigma'} \sum_{\boldsymbol{k'},\boldsymbol{G'}} \frac{v_C(\boldsymbol{k}-\boldsymbol{k'}-\boldsymbol{G'})}{N} \phi_{l,\sigma,\boldsymbol{k'}+\boldsymbol{G'}+\boldsymbol{G_m}}^{i,*} \phi_{l,\sigma ',\boldsymbol{k'}+\boldsymbol{G'}+\boldsymbol{G_n}}^{i'},
\end{equation}
where $\sigma$ and $\sigma'$ are spin indices. We considered both spin-polarized and non-polarized cases for both SS and SP solutions.  In Fig.~\ref{fig:HF_Energy} we show the total energy for these four cases. In general they are all close in energy, but we found that non-polarized solutions are favored for both SS and SP cases. Furthermore, the exchange term increases the energy of the SP solutions more than in the SS case, making the SS non-polarized solutions the ground state. The band structures do not present any significant change in shape with the inclusion of the exchange term.

\begin{figure}[ht]
\begin{center}
{\includegraphics[scale = 0.95]{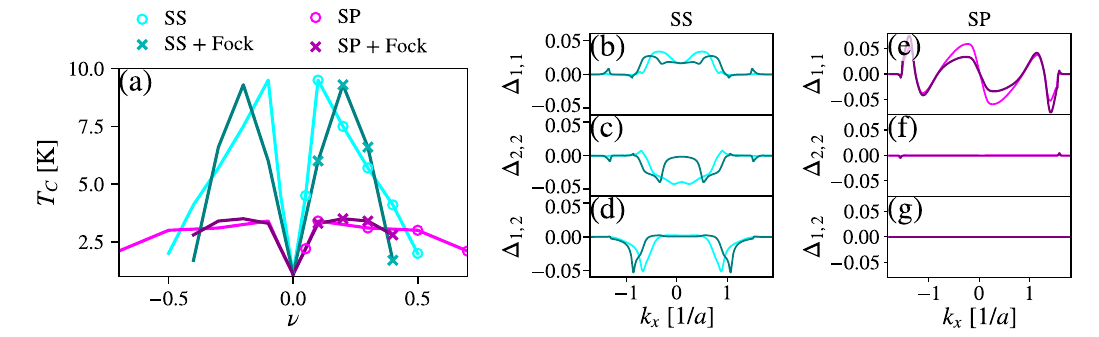}}
\caption{
Exchange term modifications to $T_C$ and OPs. (a) Comparison between critical scales with and without exchange term as function of $\nu$. A non-polarized Fock was considered as it was shown to be the ground state for both SS and SP solutions in Fig.~\ref{fig:HF_Energy}. The OPs are compared at $\nu=0.1$ (b)-(d) for SS and (e)-(g) for SP solutions.  
}
\label{fig:TC_Fock}
\end{center}
\end{figure}

We now calculate the $T_C$ and OPs with the inclusion of the exchange term for a few different values of $\nu$. In Fig.~\ref{fig:TC_Fock}(a) we compare the $T_C$ with and without exchange for both SS and SP solutions. As the non-spin-polarized solutions are favored, we only considered these in this comparison. For the SS case, there is a slight shift in the maximum $T_C$, but the overall shape and magnitudes remain within the same range. For the SP case, the $T_C$ is roughly the same as without exchange. In Figs.~\ref{fig:TC_Fock}(b)-(g) we compare the OPs at $\nu=0.1$. They remain symmetric for the SS case and antisymmetric for the SP one. For the other values of $\nu$ calculated, there is also no change in symmetries, or considerable change in magnitudes, and thus we can safely describe the pairing instability within the system without the inclusion of the exchange term.  

\section{Robustness of Superconductivity for Other Strain Parameters}

\begin{figure}[ht]
\begin{center}
{\includegraphics[scale = 0.95]{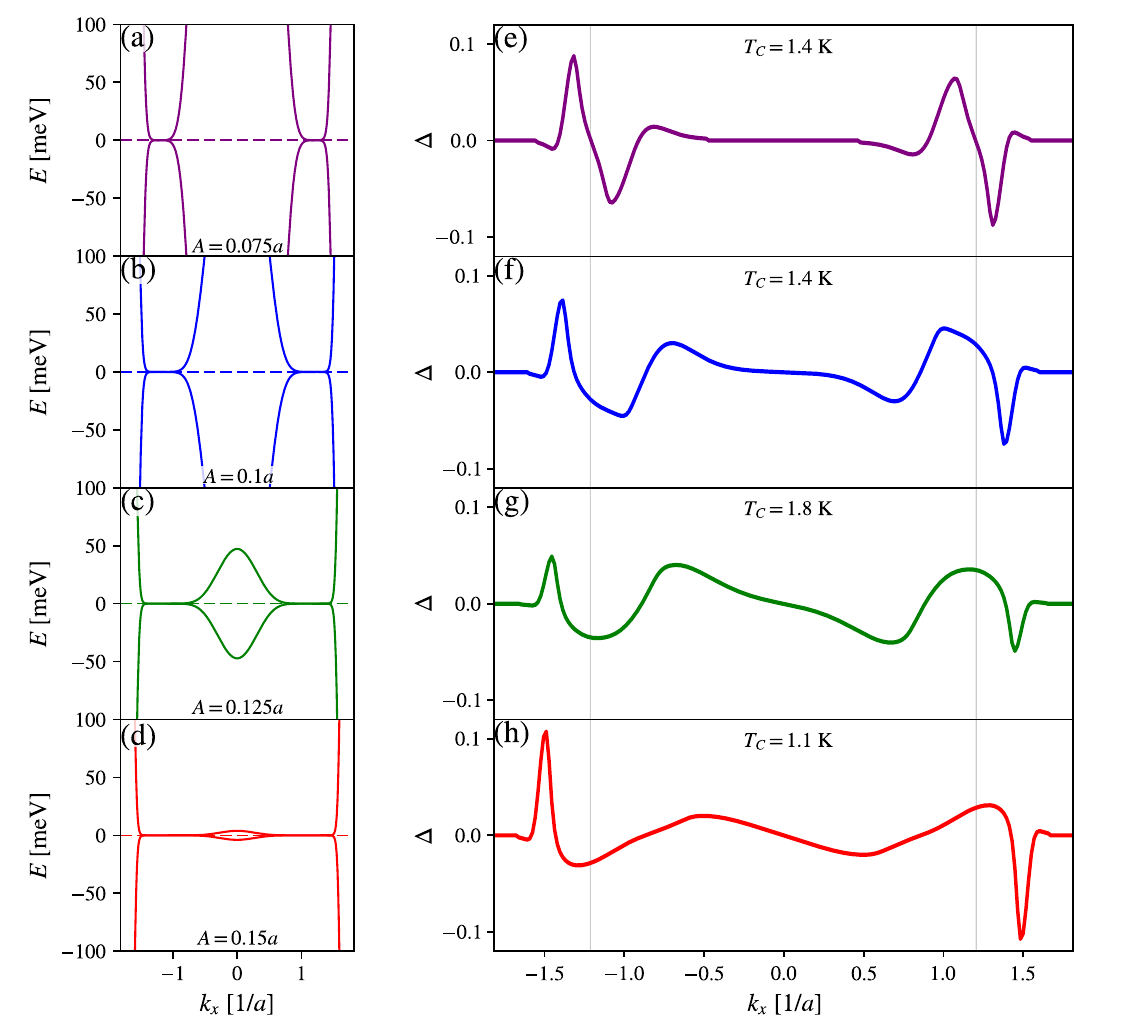}}
\caption{(a)-(d) Band structure for the NH case with $\lambda=13.6$ nm and different amplitudes of $A$. In each case, the dashed horizontal lines indicate the Fermi level for $\nu=0$, at which the $T_C$ and intraband OP ($\Delta_{1,1}$=$\Delta_{2,2}$) calculated with $\varepsilon=10$ are shown in (e)-(h). The vertical lines indicate the position of the Fermi surface. 
}
\label{fig:VarStrNH}
\end{center}
\end{figure}

We now show how the bands, order parameters, and critical scales are modified by other strain parameters such as $A$ and $\lambda$. In Fig.~\ref{fig:VarStrNH} we show the band structures and OPs at their respective $T_C$ for different values of $A$ at $\nu=0$. As the strain increases, the bands flatten over a broader range of $k_x$. Surprisingly, the highest strain value $A=0.15a$, yields the lowest $T_C=1.1$ K, while the maximum $T_C=1.8$ K is found at $A=0.125a$. However, for smaller values of $A$ the $T_C$ is expected to decrease at smaller values of $\nu$, as we exit the flat band regime earlier. A correlation can be seen between the peaks of the OPs and the values of $k_x$ at which the bands cease to be flat, this is especially clear for $A=0.075a$. As the bands flatten further, the valleys begin to merge and only the peaks towards the BZ borders remain. 

\begin{figure}[ht]
\begin{center}
{\includegraphics[scale = 0.95]{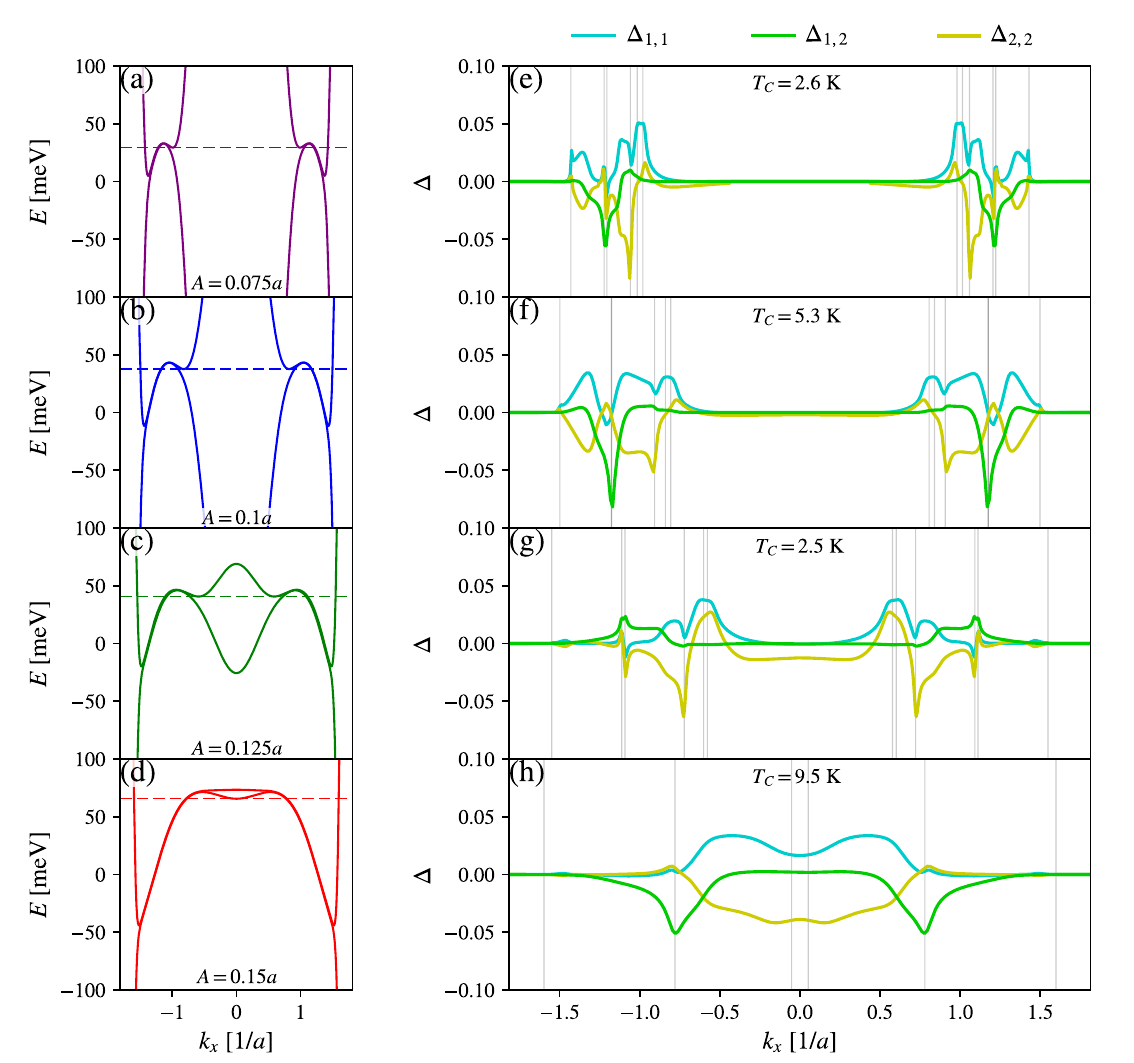}}
\caption{(a)-(d) Band structure for $\lambda=13.6$ nm and different amplitudes of $A$, including the SS Hartree potential at $\nu=0.1$ with $\varepsilon=10$. (a) $A=0.075a$, (b) $A=0.1a$, (c) $A=0.125a$ and (d) $A=0.15a$. The dashed horizontal lines indicate the Fermi level. In (e)-(h), we show for each case the $T_C$ and the OP projected onto the intraband processes of both bands ($\Delta_{1,1}$ and $\Delta_{2,2}$), as well as interband processes ($\Delta_{1,2}$).  The vertical lines indicate the position of the Fermi surface.}
\label{fig:varStrSS}
\end{center}
\end{figure}

In Fig.~\ref{fig:varStrSS} we show similarly the band structures and OPs at their corresponding $T_C$ for different values of $A$, this time for $\nu=0.1$ with the SS Hartree potential. Here the values of the $T_C$ span a wider range, ranging from 2.6~K to 9.5~K. In addition, the behavior is more intricate, as the $T_C$ depends strongly on the Fermi surface. A common feature between the two cases with the highest $T_C$ ($A=0.1a$ and $A=0.15a$) is a significant contribution from the interband terms, consistent with the analysis in the previous section. 

\begin{figure}[ht]
\begin{center}
{\includegraphics[scale = 0.95]{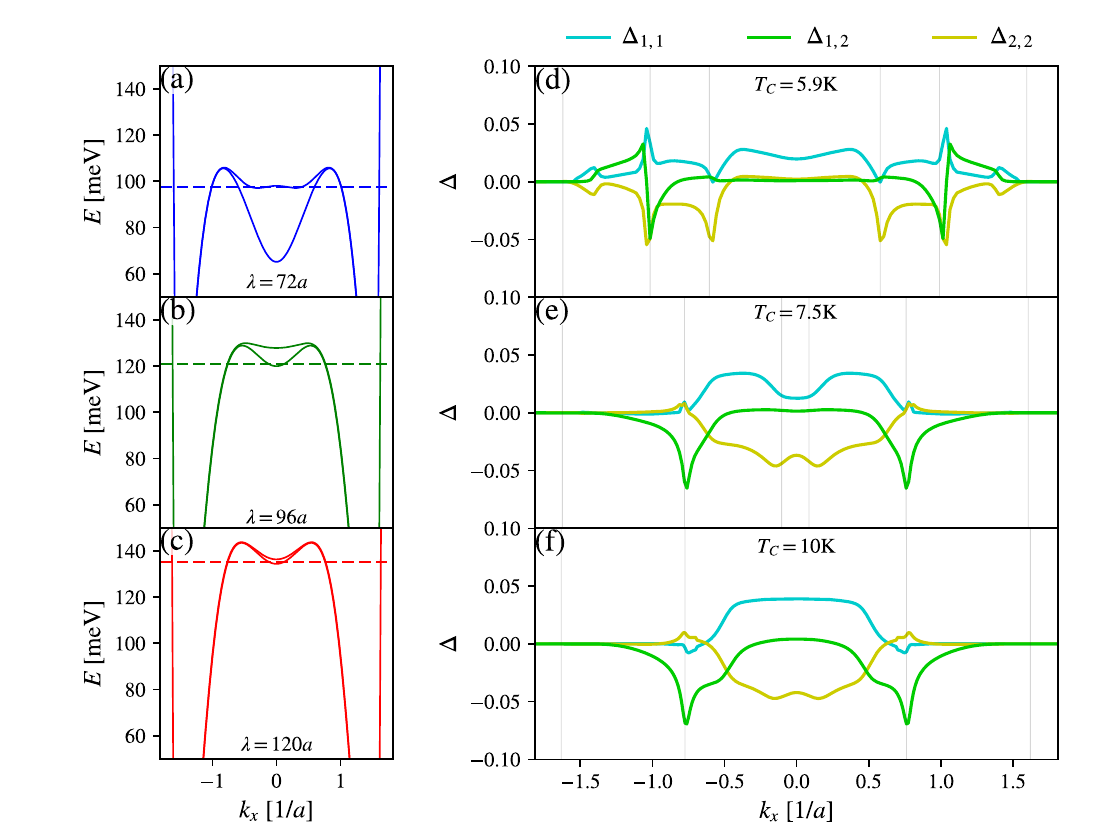}}
\caption{
(a)-(c) Band structure for $A=0.15a$ and different values of $\lambda$, including the SS Hartree potential at $\nu=0.2$ with $\varepsilon=10$. (a) $\lambda=72a$, (b) $\lambda=96a$ and (c) $\lambda=120a$. The dashed horizontal lines indicate the Fermi level. In (d)-(f), we show for each case the $T_C$ and the OP projected onto the intraband processes of both bands ($\Delta_{1,1}$ and $\Delta_{2,2}$), as well as interband processes ($\Delta_{1,2}$).  The vertical lines indicate the position of the Fermi surface.}
\label{fig:varWL}
\end{center}
\end{figure}

Similarly in Fig.~\ref{fig:varWL} we show the results for different values of $\lambda$ for the SS solution at $\nu=0.2$. As $\lambda$ defines the moir\'e length, the distance between the solitons increases with $\lambda$, reducing the hybridization between them, resulting in flatter bands. Looking at the $T_C$ values, a trend can be seen where the temperatures increase as the bands become flatter, reaching up to 10 K for $\lambda=120a$. In each case, there is a significant contribution of the interband pairing, supporting our claim on its relevance to increase the $T_C$. In general, the pairing instability persists with $T_C$ values of a few kelvin, even as the strain amplitude or wavelength are modified, indicating that precise fine-tuning of these parameters is not necessary.

\section{Screening Dependence of Critical Temperatures}
\begin{figure}[ht]
\begin{center}
{\includegraphics[scale = 0.95]{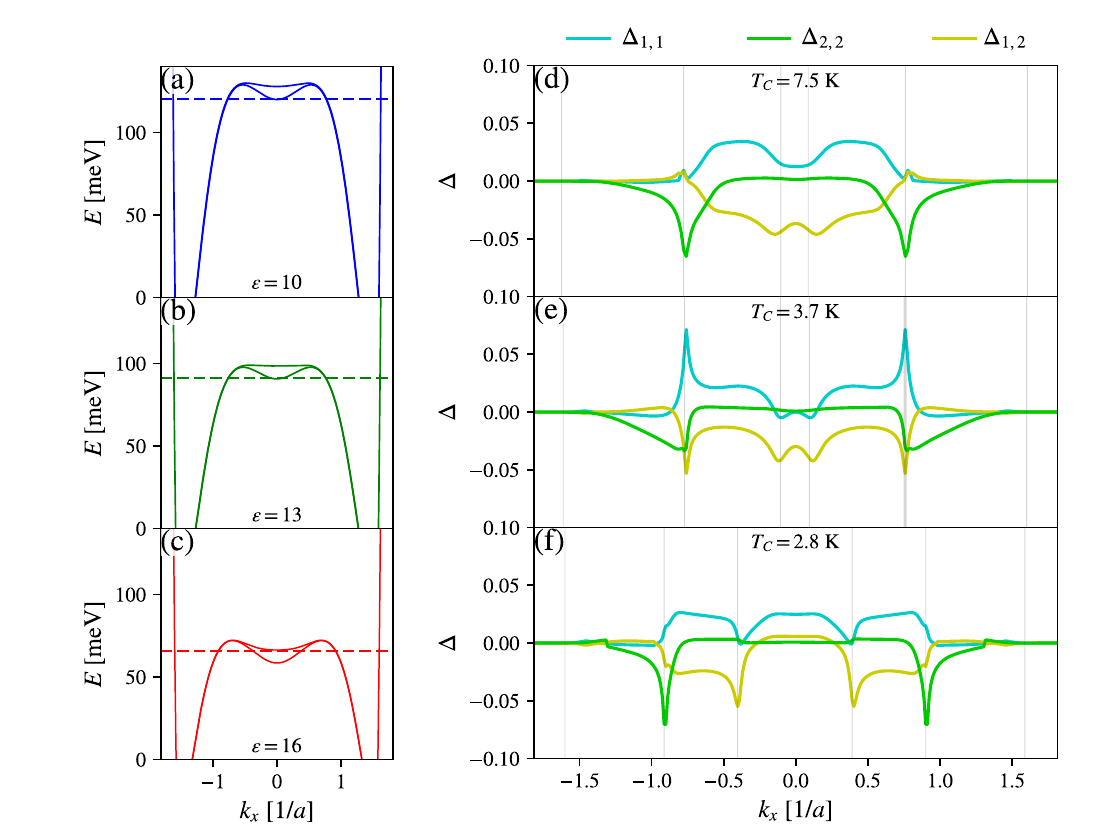}}
\caption{
(a)-(c) Band structure for the SS solution with $A=0.15a$, $\lambda=13.6$ nm and different values of $\varepsilon$. (a) $\varepsilon=10$, (b) $\varepsilon=13$ and (c) $\varepsilon=16$. The dashed horizontal lines indicate the Fermi level. In (d)-(f), we show for each case the $T_C$ and the OP projected onto the intraband processes of both bands ($\Delta_{1,1}$ and $\Delta_{2,2}$), as well as interband processes ($\Delta_{1,2}$).  The vertical lines indicate the position of the Fermi surface.
\label{fig:vareps}}
\end{center}
\end{figure}

To further test the robustness of the pairing instability, we now show how modifying the screening parameters, $\varepsilon$ and $d_g$, affects our results. In Fig.~\ref{fig:vareps} we show the bands and OPs for three different values of $\varepsilon$. As the dielectric constant increases, the Hartree potential becomes more screened, resulting in a smaller energy shift in the bands as shown in Figs.~\ref{fig:vareps}(a)-(c). For each case the OPs are found to the right, they exhibit the same symmetries and a significant interband term. As the screening increases, the interaction between electrons becomes weaker, resulting in lower values of $T_C$.

\begin{figure}[ht]
\begin{center}
{\includegraphics[scale = 0.95]{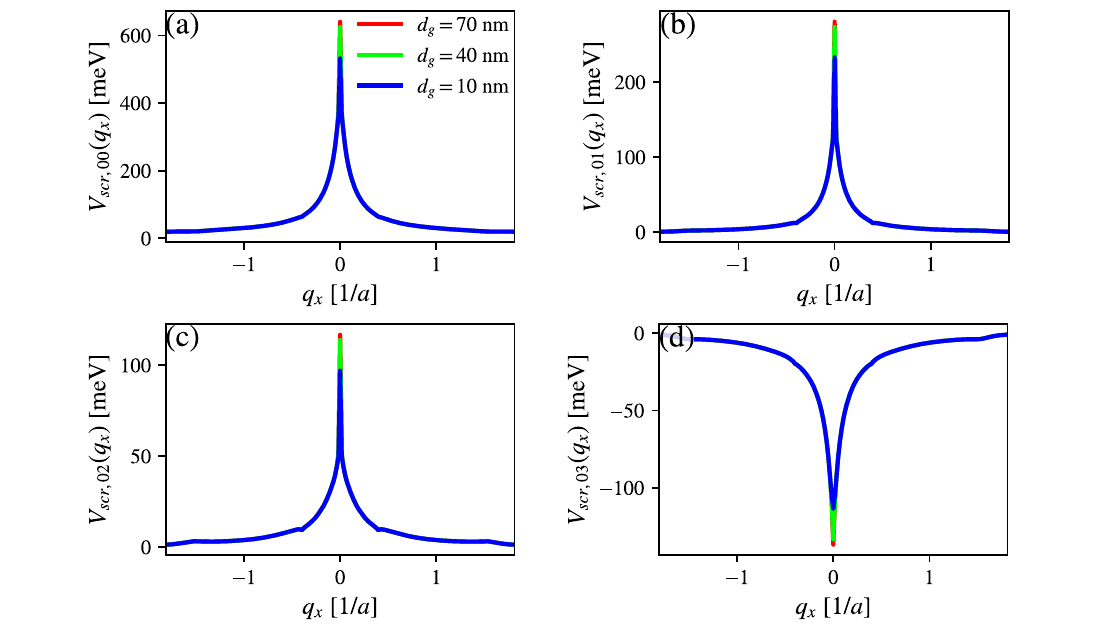}}
\caption{
Matrix elements of the screened Coulomb potential in momentum space for $A=0.15a$, $\lambda=13.6$ nm, $\varepsilon=10$ and different values of $d_g$, 70 nm (red), 40 nm (green) and 10 nm (blue). (a) ($G_0,G_0$), (b) ($G_0,G_1$), (c) ($G_0,G_2$) and (d) ($G_0,G_3$).
}
\label{fig:vardg}
\end{center}
\end{figure}

On the other hand, $d_g$ does not affect the values of the $T_C$ as it only enters the calculation through Eq. (\ref{eq:vc(q)}), which regulates the Coulomb potential at $|\boldsymbol{q}|=0$, but otherwise saturates quickly. In Figs.~\ref{fig:vardg} we show some components of the screened Coulomb potential in momentum space for different values of $d_g$. Although we considered differences of 30 nm, the shapes remain exactly the same, with just a slight variation of the magnitude at $|\boldsymbol{q}|=0$, thus variations of $d_g$ are not expected to qualitatively affect the $T_C$ or pairings.

\end{document}